\documentclass[smallextended,twocolumn]{svjour3}       

\smartqed  
\usepackage{graphicx}
\usepackage{amsmath}
\usepackage{url}
\usepackage[hang,small,bf]{caption}
\usepackage[subrefformat=parens]{subcaption}
\usepackage{color}
\usepackage[numbers,square]{natbib}
\usepackage{multirow}

\usepackage{enumitem}
\setlist[itemize]{nosep,noitemsep}
\setlist[description]{nosep,noitemsep}
\setlist[enumerate]{nosep,noitemsep, font=\bf}
\usepackage{subfloat}

\makeatletter
\let\cl@chapter\@empty
\makeatother
\usepackage[capitalise]{cleveref}
\crefname{theo}{Theorem}{Theorems}
\crefname{prop}{Proposition}{Proposition}
\crefname{lemma}{Lemma}{Lemma}
\crefname{eq}{Eq.}{Eqs.}
\crefname{fig}{Fig.}{Figs.}
\crefname{tab}{Tab.}{Tabs.}
\crefname{sec}{Section}{Sections}

\def\({\left(}
\def\){\right)}
\def\<<{\left\{}
\def\>>{\right\}}

\AtBeginDocument{\setlength{\mathindent}{2em}}

\makeatletter
\def\makeheadbox{}
\makeatother

\begin{document}

\title{Estimation of jam density and saturation flow at signalized intersections using only low penetration rate connected vehicle GNSS trajectories: Case study using actual data in Tokyo}

\titlerunning{Jam density and saturation flow  estimation}        

\author{Toru Seo         \and
        Takahiko Kusakabe \and
        Takashi Oguchi
}


\institute{T. Seo \at
				Institute of Science Tokyo\\
				2-12-1-W6-10 Ookayama, Meguro, Tokyo 152-8552, Japan\\
				\email{seo.t.aa@m.titech.ac.jp}           
           \and
           T. Kusakabe \at
           		Social Designers Base\\
				\email{t.kusakabe@sd-base.jp}           
           \and
           T. Oguchi \at
           		Institute of Industrial Science, the University of Tokyo\\
				\email{takog@iis.u-tokyo.ac.jp}           
}

\date{Preprint (``Submitted Manuscript'') accepted by International Journal of Intelligent Transportation Systems Research.\\ First submission: 2026-05-29. Accepted: 2026-10-16}

\journalname{}
\maketitle

\begin{abstract}
	The jam density and saturation flow at a signalized intersection are basic inputs to traffic signal management schemes that use connected vehicle (CV) trajectory data.
	However, methods for estimating the jam density itself from low penetration rate CV trajectories have received little attention.
	Although a few theoretical methods have been proposed, their application is very limited and it is not clear whether those parameters can be estimated using actual CV data, especially with low penetration rates.
	This paper analyze such methods, in which the jam density and saturation flow are estimated from long-term historical distribution of low penetration rate CV stop positions accumulated over many signal cycles.
	The methods are evaluated using actual CV trajectory data collected at three signalized intersections in Tokyo for six months, with a penetration rate on the order of 1--3{\%}.
	The estimators mostly agree on a jam spacing close to 8~m at the sites at which the queueing pattern is well developed, and the saturation flow estimates lie within roughly 10{\%} of an independent video-based reference at those sites.
	Furthermore, quantitative criteria for self-evaluation of reliability of the estimates are proposed.

\end{abstract}

\section{Introduction}

The {\it saturation flow} (i.e., traffic capacity) and {\it jam density} (i.e., jam spacing) are some of the basic inputs in standard traffic signal control.
Recent studies have found that their actual values vary non-trivially across sites and over time depending on vehicle composition, road geometry, weather, and driver behavior \cite{Aoyama2020capacity,Knoop2022capacity}.
Traditional measurement of these values, however, relies on fixed observations such as inductive loop detectors or manual video surveys, both of which are costly to deploy and maintain at the scale of an urban road network.

{\it Connected vehicle} ({\it CV}; also referred to as probe vehicle) data have attracted attention as an alternative source of traffic information \cite{Seo2017review, Guo2019signal}.
In this paper, the term ``CV'' refers to vehicles that are randomly present in the traffic flow and continuously report their positions using onboard Global Navigation Satellite System (GNSS).
Various studies have proposed methods to estimate the saturation flow or the fundamental diagram (FD) at signalized intersections from CV data \cite{Herrera2010probe, seo2019probefd, Bagheri2015signal, Tan2024signal, Wang2024signal, Seo2026signalcv_en}, and complementary lines of work have addressed related estimation problems such as queue length \cite{Comert2013signalintersection}, arrival volume \cite{wada2015probe, Zheng2017intersection}, individual vehicle trajectories \cite{Sun2013signalintersection}, and the CV penetration rate itself \cite{Wong2019probe}.
Empirical validation of these saturation flow estimation methods using actual CV data has also been widely studied \citep[e.g.][]{Wang2024signal}.

However, most of these saturation flow estimation methods rely on the value of the jam density as an exogenous assumption or external input.
This assumption does not always hold, because the jam density depends on vehicle composition, road geometry, and driver behavior and is rarely known for an arbitrary intersection.
A misspecified jam density introduces undesirable bias into the estimated capacity.

In contrast, dedicated estimation of the jam density itself from low penetration rate CV data has received little attention.
Lloret-Batlle and Zheng \cite{LloretBatlle2023signalcv} have proposed methods to estimate the jam density using the stop positions of CVs over multiple cycles.\footnotemark{}
The CV penetration rate was assumed to range from 5{\%} to 25{\%}, and their methods were evaluated using simulation and NGSIM data.
Wang et al.~\cite{Wang2019tse} have proposed a joint FD parameter estimation method using a Bayesian network that combines shockwave analysis with prior assumptions on parameter distributions; however, they reported only the overall performance of the joint estimator and did not present a per-parameter analysis for the jam density alone.
Their evaluation is also based on simulation and NGSIM data.
Therefore, the achievable accuracy of jam density estimation from actual long-term CV data, and the extent to which different estimators agree on it, have remained open questions.

\footnotetext{
	Although Lloret-Batlle and Zheng \cite{LloretBatlle2023signalcv} state in their abstract that ``We use as measurements the distances between successive vehicle stops,'' (which implies that multiple CVs must stop consecutively within the same cycle), some of their mathematical formulations actually allow for using the absolute stop position of a single CV per cycle.
	This distinction is crucial for sparse data environments: while the former requirement demands a high penetration rate to capture consecutive CVs, the latter interpretation functions effectively even at a low penetration rate where observing more than one CV in a single queue is rare.
	This single-CV interpretation can be considered as a stronger and more general measurement setting than the one stated in the abstract, and this paper adopts this interpretation.
}

Another research field that remains relatively unexplored is the very low penetration rate regime.
The majority of CV-based signal estimation and/or control studies assume penetration rates of around 10{\%} or higher \cite{Zheng2017intersection, Tan2024signal, Wang2024signal}; only a small number of studies has explicitly targeted lower regimes.
The work of \cite{feng2018signal} proposed a detector-free adaptive signal control method valid down to a 2{\%} penetration rate, and \cite{day2016signal} examined an offset-optimization scheme at 1{\%} penetration with a suggested extension to 0.1{\%} through multi-day data pooling.

The combination of jam density estimation and very low penetration rate CV data presents unique challenges.
Existing jam density estimation methods \cite{Wang2019tse,LloretBatlle2023signalcv} assume that they use long-term, historical CV data over multiple cycles and estimate the jam density by analyzing statistical patterns in the CV data distribution.
However, it is not clear whether such statistical patterns actually exist in such a way that the jam density can be estimated.
The CV data distribution should depend on various conditions such as traffic characteristics, location characteristics, and data collection techniques that cannot be simulated.
Although validation using simulation or one-day data like NGSIM, as used by \cite{Wang2019tse,LloretBatlle2023signalcv}, is useful for proof-of-concept of theoretical methods, it does not clarify the actual performance of such methods in low penetration rate CV environments.

This paper fills this gap.
Specifically, this study proposes and empirically validates methods for estimating the jam density and saturation flow at signalized intersections from low penetration rate CV trajectory data.
The proposed methods exploit the long-run statistical pattern of CV stop positions accumulated over many signal cycles.
Although the fundamental idea of the proposed methods is essentially the same as that in the existing works by Lloret-Batlle and Zheng \cite{LloretBatlle2023signalcv} and Seo et al.~\cite{Seo2026signalcv_en}, we propose suitable implementations for actual CV data.
Then, the proposed methods are validated using actual long-term CV data---six months of ITS Connect \cite{ITS2019probe_en} trajectory data collected at three signalized intersections in Tokyo---with reference values obtained from video observation.
We show both positive and negative results in order to obtain insights into the applicable range of the proposed methods.

The remainder of this paper is organized as follows.
Section 2 describes the proposed jam density and saturation flow estimation methods.
Section 3 reports the empirical validation results.
Section 4 concludes this paper.

\section{Methods}

\subsection{Overview}

The inputs for the proposed methods are
\begin{itemize}
	\item High-resolution, map-matched CV trajectory data on an approach to a signalized intersection.
	\item Signal timing information for that approach.
\end{itemize}
for a sufficiently long period.
The CV penetration rate and traffic flow model parameters (e.g., free-flow speed, wave speed, jam density, vehicle size) are unknown.

\begin{figure}[tbp]
	\centering
	\subfloat[Jam spacing estimation]{\includegraphics[clip, width=0.8\hsize]{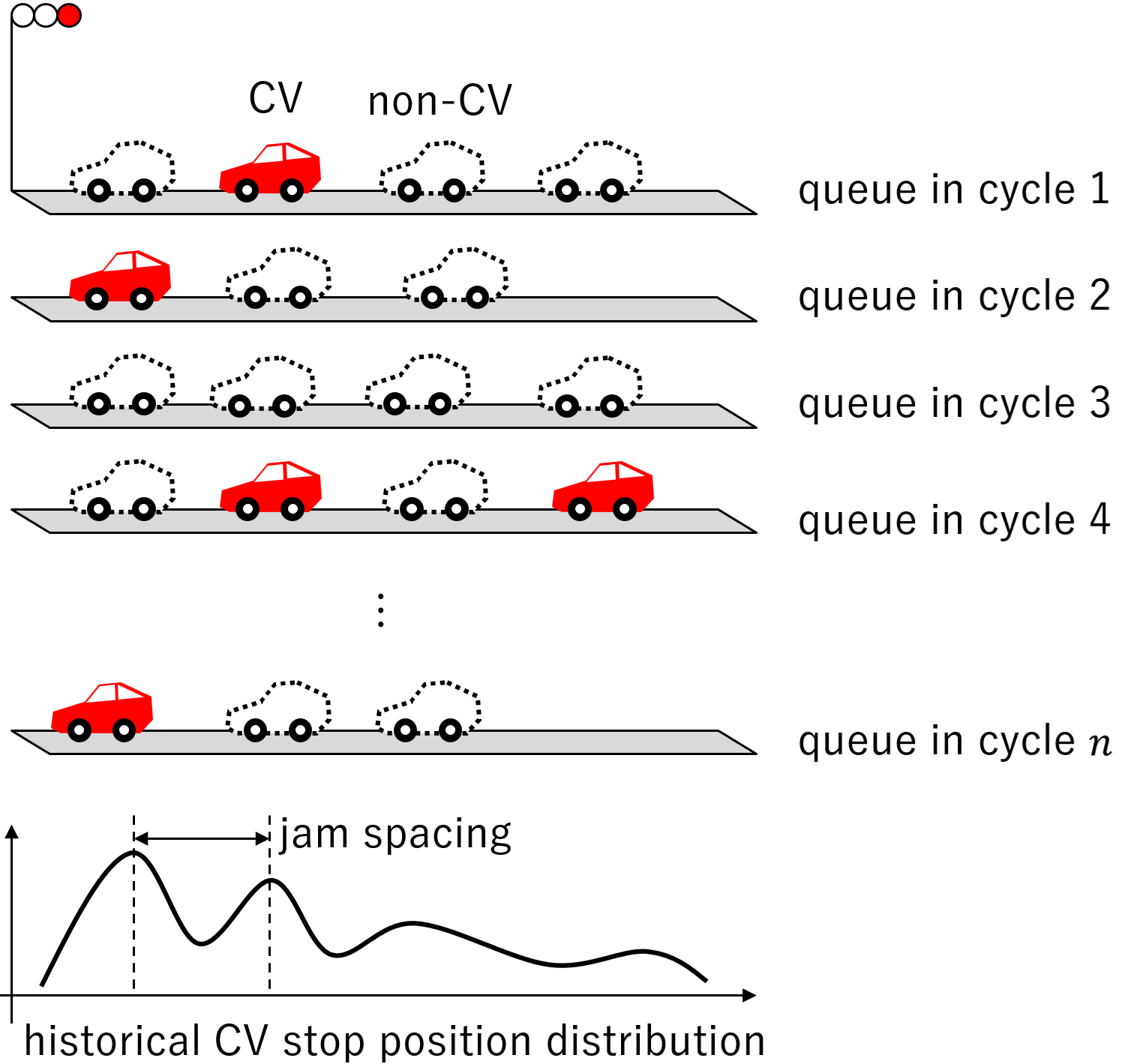}\label{concept_jam_spacing}}\\\vspace{1em}
	\subfloat[Saturation flow estimation: pass time (adopted from \cite{Seo2026signalcv_en})]{\includegraphics[clip, width=0.99\hsize]{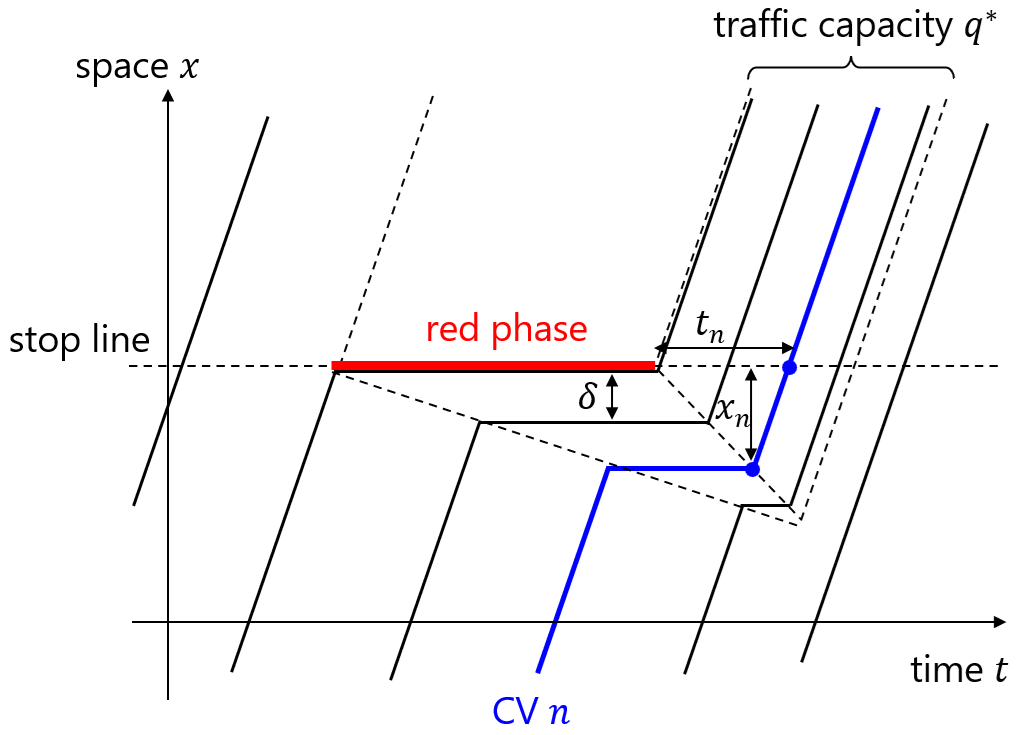}\label{concept_capacity}}\\\vspace{1em}
	\subfloat[Saturation flow estimation: regression (adopted from \cite{Seo2026signalcv_en})]{\includegraphics[clip, width=0.99\hsize]{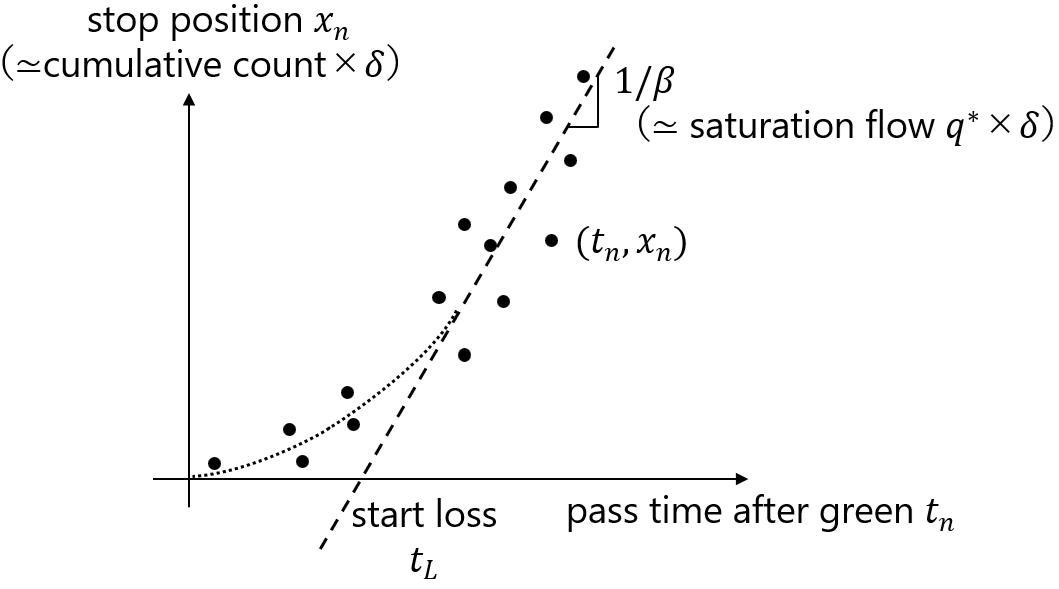}\label{concept_capacity2}}
	\caption{Conceptual illustration of the basic idea of the proposed method.}
	\label{concept}
\end{figure}

The proposed method estimates the jam density first.
The basic idea is illustrated in \cref{concept_jam_spacing}.
Typically, the first vehicle in a signalized intersection's waiting queue stops at a certain location determined by the stop line.
The second vehicle stops while keeping the jam spacing distance $\delta$ from the first vehicle.
The subsequent vehicles do likewise.
By using CVs, we can observe the stop position of a random vehicle in some cycles.
The stop position of these CVs over multiple cycles varies due to heterogeneity and GNSS) measurement error.
However, when CVs' stop positions are accumulated over a sufficiently long observation period, the resulting empirical distribution is expected to exhibit a regular peaked pattern, with peaks located at multiples of the average jam spacing measured from the head of the queue.
By measuring the distance between peaks, we can estimate the value of the jam spacing.

\begin{figure*}[tbp]
	\centering
	\subfloat[Time-space trajectory plot]{\includegraphics[clip, width=0.8\hsize]{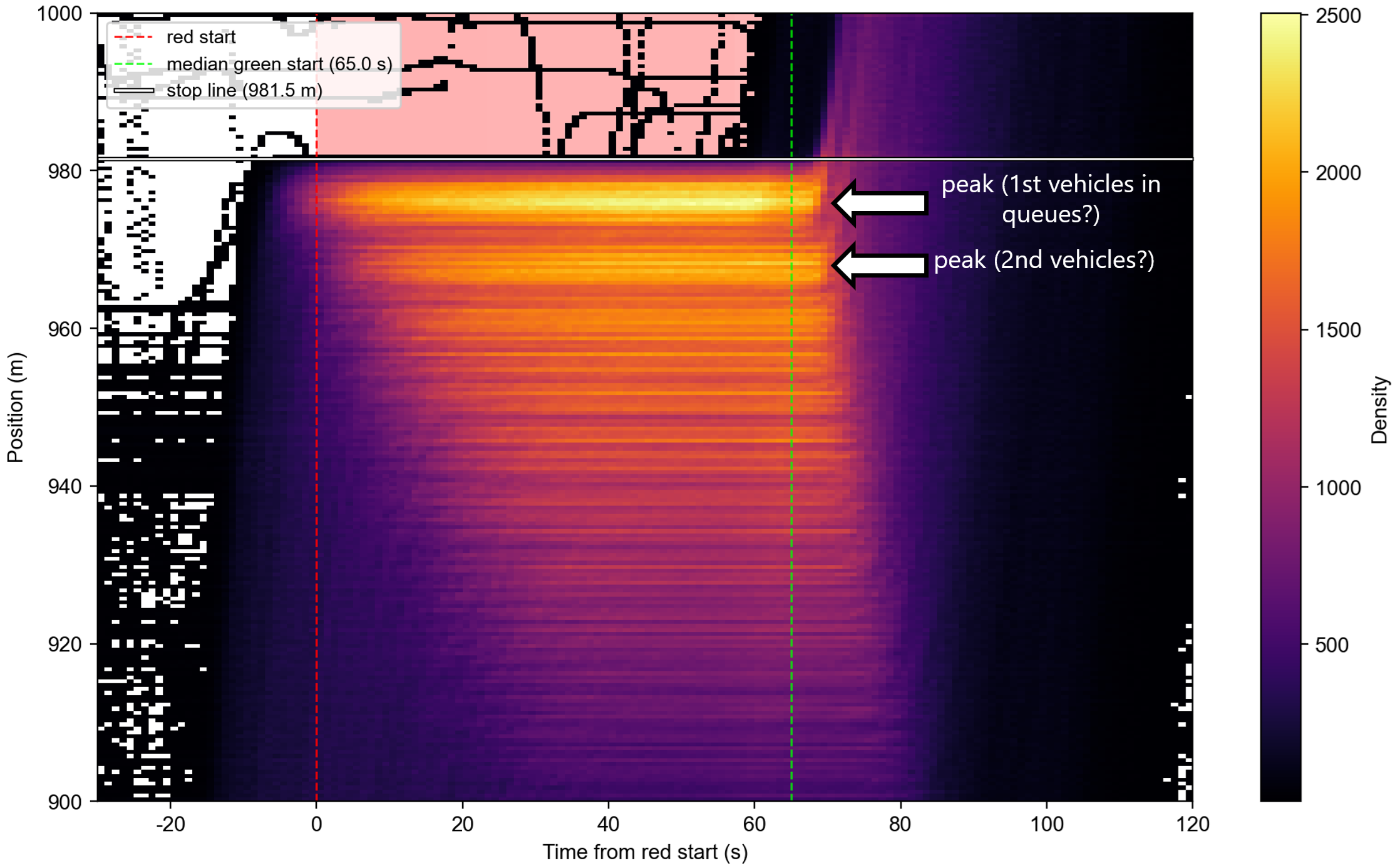}\label{stop_distribution_tsd}}\\
	\subfloat[Stop position distribution]{\includegraphics[clip, width=0.6\hsize]{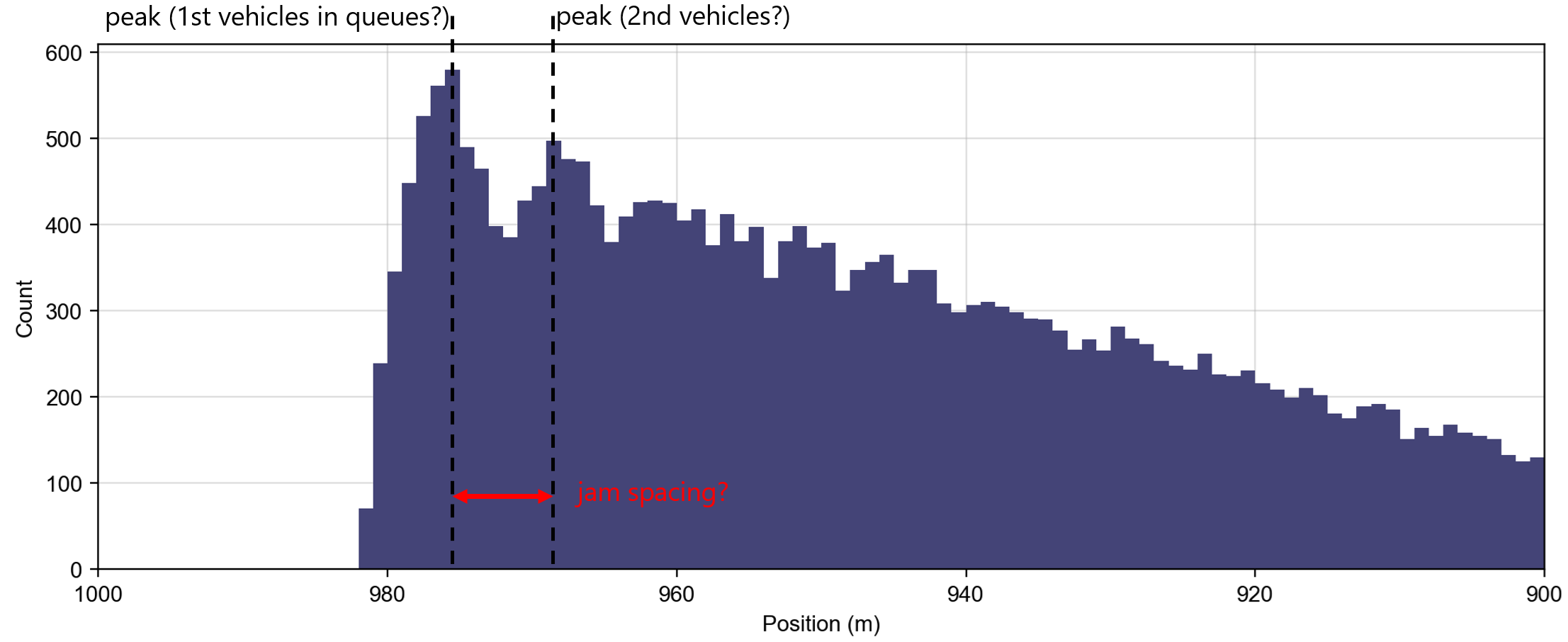}\label{stop_distribution_hist}}
	\caption{Per-cycle CV trajectory distribution at an intersection over half a year in Tokyo.}
	\label{stop_distribution}
\end{figure*}

An actual CV stop position distribution obtained from one of the test sites in this study is shown in \cref{stop_distribution}.
It is derived from 2{\%} penetration rate CV data over six months; for the detailed specification, see \cref{sec:data}.
The top panel shows CV trajectories over multiple cycles, synchronized by the red phase start time, as a time--space diagram, and the lower panel shows the distribution of stop positions.
We can see that this distribution has clear peaks around 970~m and 960~m, followed by relatively unclear peaks separated by intervals of several meters.
This pattern is consistent with the conceptual illustration in \cref{concept}, and it can be inferred that the first peak corresponds to the first vehicle in the queue, and so on.

After obtaining the jam spacing estimate, the saturation flow can be estimated by using the method of Seo et al.~\cite{Seo2026signalcv_en}.
The basic idea of this method is illustrated in \cref{concept_capacity,concept_capacity2}.
In this method, the cumulative departure curve is reconstructed from the pass times of CVs over multiple cycles.
The cumulative count can be estimated by dividing the stop position $x_n$ by the estimated jam spacing $\delta$.
Then, performing a certain regression, we can obtain the value of the saturation flow $q^*$ (\cref{concept_capacity2}).

Both stages require a sufficiently long observation period due to the following reasons.
For the jam density estimation, individual stop positions are noisy owing to vehicle heterogeneity and GNSS error, and only the aggregated distribution exhibits a meaningful peaked pattern.
For the saturation flow estimation, the regression must be supported by a sufficient number of $(x_n, t_n)$ pairs spread across the queued region.
The proposed method is designed for offline estimation, in which CV data are accumulated for a certain period and the estimation is performed retrospectively on the accumulated record.
The necessary data accumulation period must depend on the CV penetration rate; the lower the penetration rate, the longer the accumulation period.

\subsection{Jam density estimation methods}

Let $\{x_n\}_{n=1}^{N}$ denote the upstream stop positions of $N$ CVs that stopped on the studied approach during the observation period.
The proposed estimators all operate on a spatial histogram of $\{x_n\}$ constructed on the analysis range $[0, X_{\max}]$ with a bin width of $\Delta X$, which is fine enough to resolve peak intervals of several meters without introducing excessive empty-bin noise.
Each estimator returns an estimate $\hat{\delta}$ of the jam spacing $\delta$, defined as the mean longitudinal distance between consecutive stopped vehicles.

We propose three estimators that differ both in the portion of the histogram they exploit and in the statistical assumptions they rely on.
This diversity is by design: because they are based on different assumptions, agreement among them on a given site provides empirical support that the recovered jam spacing is a property of the underlying queue rather than an artifact of any particular procedure.
For illustration of the methods, the actual estimation results shown later (\cref{jam_res_detailed}) would be informative.

\subsubsection{Gaussian mixture method}\label{method_G}

The first method is the {\it Gaussian mixture method}, which is based on the observation that the empirical histogram near the head of the queue should resemble the superposition of two unimodal density components.
Each component is broadened by the cycle-to-cycle variability of the head-of-queue offset and by the GNSS positioning error of the CVs, both of which are reasonably approximated as Gaussian noise.

We model the histogram counts on the analysis range $[0, X_{\max}]$~m as the sum of a linear baseline and two Gaussian peaks,
\begin{align}
	f(x) = a x + b + \sum_{i=1}^{2} A_{i} \exp\!\left(-\frac{(x - \mu_i)^2}{2 \sigma_i^2}\right),
	\label{eq_two_gauss}
\end{align}
where $(a, b)$ are the baseline coefficients and $(A_i, \mu_i, \sigma_i)$ are respectively the amplitude, mean, and standard deviation of the $i$-th Gaussian component.
The eight parameters are estimated by non-linear least squares.
The mean jam spacing estimate is the difference between the fitted peak centers,
\begin{align}
	\hat{\delta}_{\text{G}} = \mu_2 - \mu_1.
\end{align}

The Gaussian mixture method can be considered the most ``normal'' estimation method, as it is based on the Gaussian error assumption.
This assumption can be a limitation of this method as well, since the actual noise in this kind of CV data is not necessarily Gaussian.

\subsubsection{Autocorrelation method}\label{method_A}

The second is the {\it autocorrelation method}.
It exploits the fact that, if the histogram is approximately periodic with period $\delta$, then it should be similar to itself under a translation by $\delta$ along the queue.
The natural statistical formalization of this self-similarity is the spatial autocorrelation function of the histogram counts.
The jam spacing can therefore be read off as the lag at which the first non-trivial peak of the autocorrelation occurs.

We linearly detrend the histogram counts on the analysis range $[0, X_{\max}]$~m, smooth them by a Gaussian filter with a standard deviation of $1$~m, and compute the normalized autocorrelation $\rho(\ell)$ where $\ell$ is the lag size.
Because $\rho(\ell)$ trivially attains its global maximum at $\ell = 0$, we discard the main lobe by locating the first zero-crossing $\ell_0$ of $\rho(\ell)$ and searching only at lags beyond it.
The jam spacing estimate is then the lag at the first secondary peak,
\begin{align}
	\hat{\delta}_{\text{A}} = \arg\!\max_{\ell > \ell_{0}} \rho(\ell).
\end{align}

The Autocorrelation method pools information from every queue position into a single summary statistic and is therefore expected to be more robust than estimators that rely on the geometry of individual peaks.
Because it operates in the spatial domain and imposes no parametric model on the histogram shape, it is also expected to be relatively insensitive to deviations of the periodic pattern from any specific functional form.

\subsubsection{Rule-based method}\label{method_R}

The third one is the {\it rule-based method}, which directly locates the histogram's first two peaks by a simple rule.
If the conceptual picture of \cref{concept} is approximately correct, then these two peaks correspond to the first and second queueing positions of the average cycle, and their separation is by construction the jam spacing of the first queueing pair.
The method makes no further use of the histogram and therefore relies only on the well-behavedness of the first two peaks.

The histogram on the analysis range $[0, X_{\max}]$~m is smoothed by a Gaussian filter with a standard deviation of $1.0$~m, and local maxima are extracted using a standard peak-finding rule.
The rule applies a minimum height threshold equal to $20{\%}$ of the mean of the smoothed histogram, together with a minimum inter-peak distance of $2$~m, both of which serve to suppress spurious local maxima driven by sampling noise.
Denoting the positions of the first two detected peaks ordered by distance from the stop line by $x_{p_1}$ and $x_{p_2}$, the jam spacing estimate is
\begin{align}
	\hat{\delta}_{\text{R}} = x_{p_2} - x_{p_1}.
\end{align}

The advantage of the Rule-based method is its simplicity: it requires no parametric assumption about the histogram shape.
However, the same simplicity is its main weakness, as it discards all other information in the data, and if the peaks are not significant, it may produce erroneous results.

\subsection{Saturation flow estimation method}\label{sec_capacity}

Once the jam spacing $\delta$ has been estimated by any of the three methods of \cref{method_G,method_A,method_R}, the saturation flow  $q^{*}$ is obtained by the regression-based capacity estimator of \cite{Seo2026signalcv_en}.
The basic idea is illustrated in \cref{concept_capacity,concept_capacity2}.
For each CV $n$ that stopped on the studied approach, let $x_n$ denote the stop position measured upstream from the stop line, and let $t_n$ denote the time elapsed from green onset until CV $n$ crossed the stop line in the same cycle.
By dividing $x_n$ by $\delta$, we can estimate the CV's vehicle count in the corresponding queue, $N_n$.
Then, we can reconstruct the cumulative departure count at the stop line by accumulating $(N_n, t_n)$ over multiple cycles.
By performing a regression on this cumulative count, the saturation flow $q^*$ can be estimated.

Formally, \cite{Seo2026signalcv_en} models the expected $(x_n, t_n)$ relationship as the linear regression
\begin{align}
	t_n = \beta\, x_n + t_L,
	\label{eq_qreg}
\end{align}
where $\beta$ is the slope and $t_L$ the startup loss intercept (\cref{concept_capacity2}).
The slope is related to the saturation flow and the jam spacing by $\beta = \delta / q^{*}$, so the saturation flow estimate is
\begin{align}
	\hat{q}^{*} = \frac{1}{\hat{\beta}\, \delta}.
	\label{eq_qstar}
\end{align}
We estimate $(\hat{\beta}, \hat{t}_L)$ by quantile regression \cite{seabold2010statsmodels} at the median, which is robust against the heavy-tailed errors typically present in CV-based $(x_n, t_n)$ data.
Records with $x_n$ below $X_{\mathrm{minsat}}$~m are excluded because they are dominated by the startup-loss region, records with $x_n$ above $X_{\mathrm{maxsat}}$~m are excluded as stops unrelated to the studied signal, and records with $t_n$ above $T_{\mathrm{maxsat}}$~s are discarded as data artifacts.

\subsection{Error estimation}\label{sec_bootstrap}

The estimation uncertainty in both $\hat{\delta}$ and $\hat{q}^{*}$ is quantified by a standard nonparametric bootstrap method \cite{efron1994bootstrap}.
For each site, $B = 500$ replicates are drawn by resampling the per-CV records with replacement, where each record carries both the stop position $x_n$ and the green-onset-to-crossing time $t_n$.
On each replicate, all three jam spacing estimators of \cref{method_G,method_A,method_R} and the quantile regression for $(\beta, t_L)$ in \cref{eq_qreg} are recomputed jointly on the resampled record set, and the saturation flow on a replicate is then evaluated as $q^{*}_{(i)} = 1 / (\beta_{(i)} \delta_{(i)})$ for each method $i \in \{\text{G}, \text{A}, \text{R}\}$.


%
%
%
%

\section{Empirical validation}

\subsection{Data}\label{sec:data}

In order to conduct empirical validation of the proposed method under realistic conditions, we use actual CV data.
We used a CV dataset obtained from vehicles equipped with the ``ITS Connect'' system \cite{ITS2019probe_en}, one of the standard CV specifications in Japan.
This data records time, latitude and longitude measured by GNSS, vehicle state, and vehicle attributes at approximately 100 ms intervals, and is collected through vehicle-to-infrastructure communication with roadside units installed at intersections.
The data were collected and provided by UTMS Society of Japan.
The data collection period for this particular dataset was weekdays from 2024-12-01 to 2025-05-31.

The following intersections and directions were selected as test sites due to their scale, traffic demand, and data collection capabilities:
\begin{itemize}
	\item Ebisu 3: north-to-south direction (NS) and south-to-north direction (SN)
	\item Shinbashi 4, NS and SN
	\item Shinjuku 7, NS and SN
\end{itemize}
They are standard signalized intersections in Japan.
Each direction had non-negligible traffic demand.
These signals were controlled in response to traffic, but the operation patterns were mostly stable, with cycle length 110--150 s and green time 60--70 s.
The lane composition is summarized in \cref{intersection_lane}.
For validation purposes, the ground-truth jam density and saturation flow at these sites were obtained by analyzing video data and aerial photographs by \cite{Seo2026signalcv_en}.
The jam spacing was estimated as 7.88~m.

\begin{table}[tb]
	\centering
	\caption{Lane composition of intersections}
	\label{intersection_lane}
	\footnotesize
	\begin{tabular}{lll}
		\hline
		Site & Lanes (from left to right) \\
		\hline
		Shinbashi 4 NS & left-or-straight/straight/straight/right \\
		Shinbashi 4 SN & left-or-straight/straight/straight/right \\
		Shinjuku 7 NS & left-or-straight/straight/right \\
		Shinjuku 7 SN & left-or-straight/straight/right \\
		Ebisu 3 NS & left-or-straight/straight/right \\
		Ebisu 3 SN & left-or-straight/straight/right \\
		\hline
	\end{tabular}
\end{table}

\begin{figure}[tbp]
	\centering
	\includegraphics[clip, width=0.99\hsize]{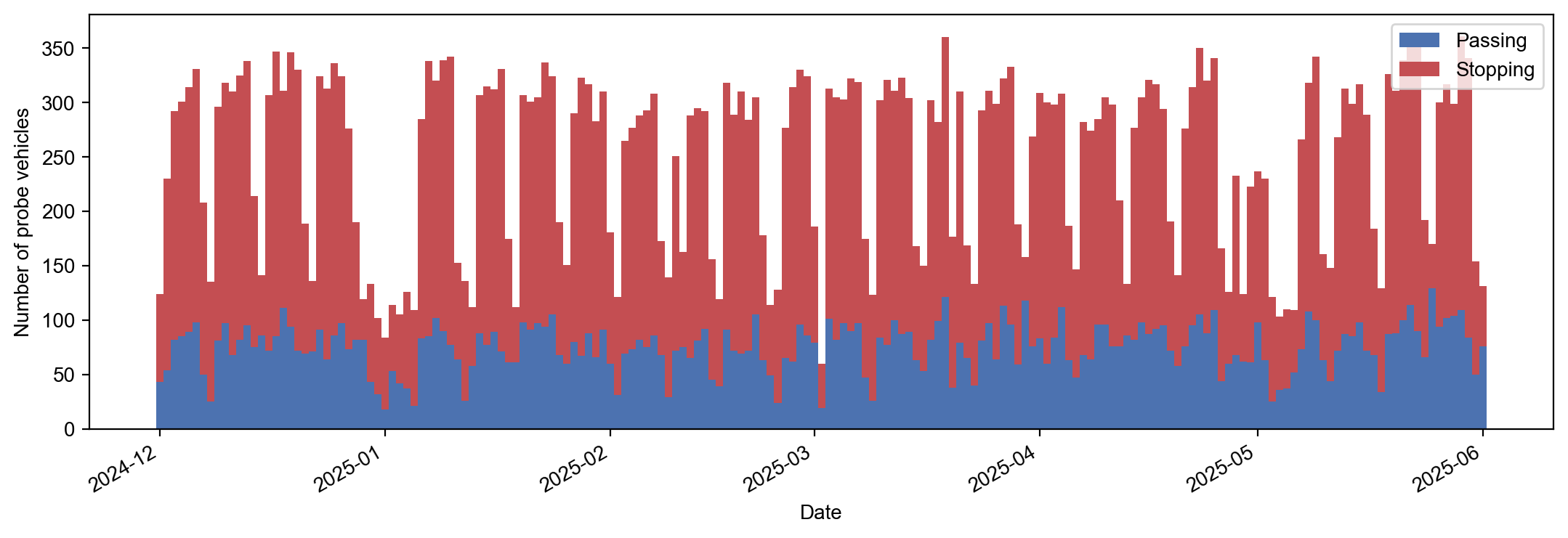}
	\includegraphics[clip, width=0.99\hsize]{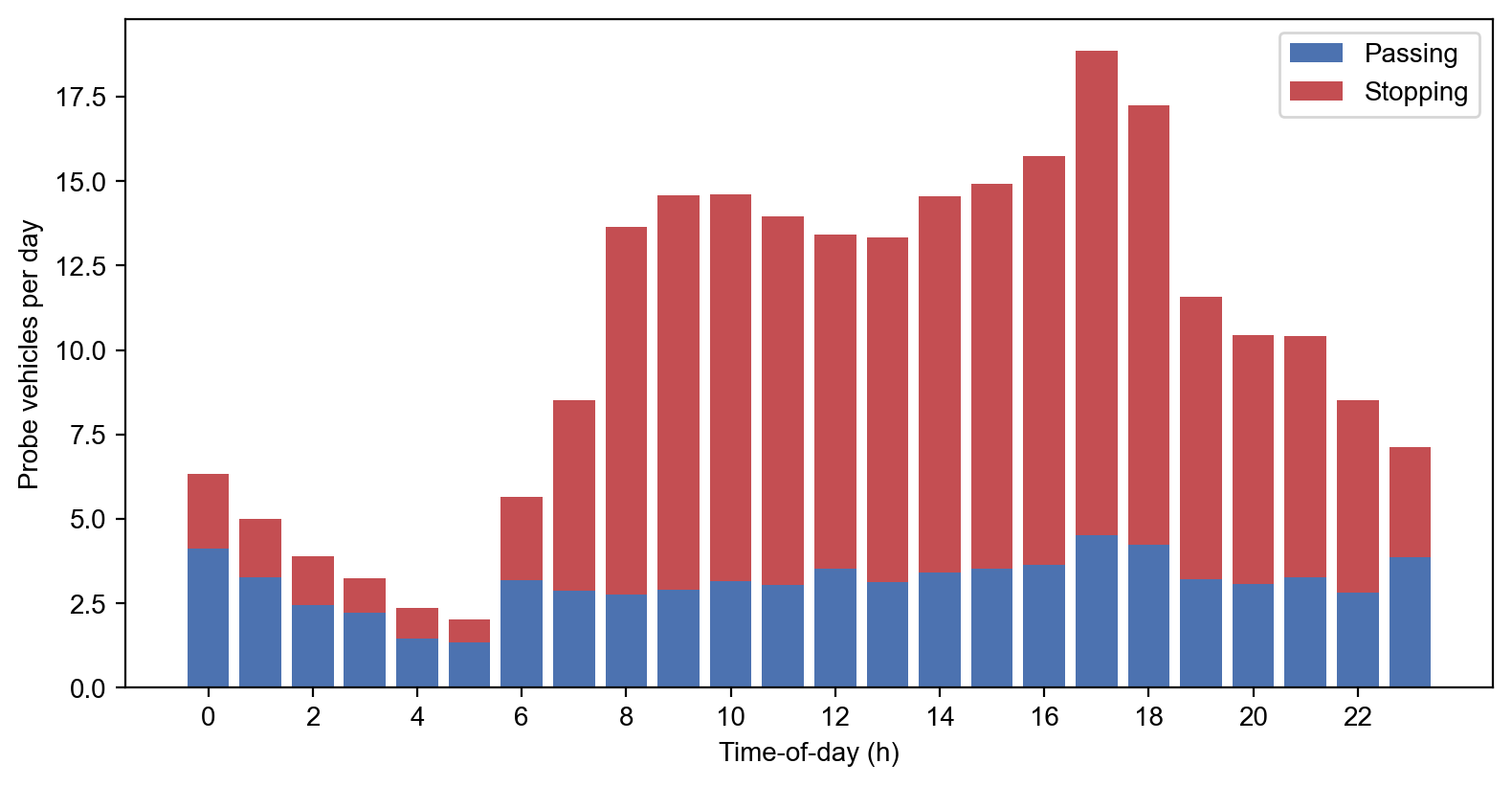}
	\caption{Number of CVs per day and time-of-day at Shinbashi 4 SN.}
	\label{number_of_CV}
\end{figure}

Basic statistics of CV data associated with the test sites are summarized in \cref{basic_stat_cv}.
The average daily traffic was measured by traffic detectors in a different period.
The estimated penetration rates of the CVs are quite small compared with those in typical CV-based estimation studies.
For a representative site (Shinbashi 4 SN), the number of CVs per day and per time-of-day is illustrated in \cref{number_of_CV}.
Typical patterns can be observed: CV counts were smaller on weekends and during the night/early morning.

\begin{table*}[tbp]
	\begin{center}
		\caption{Basic statistics of CV data per test sites for six months}
		\label{basic_stat_cv}
		\begin{tabular}{ccccccc}
			\hline
			intersection & direction & CV count & CV stopped & CV per day & Average daily traffic & CV penetration rate \\
			\hline
			Ebisu 3 & NS & 52926 & 14633 & 289.21 & 9496 & 3.0{\%} \\
			Ebisu 3 & SN & 55106 & 25139 & 301.13 & 7411 & 4.1{\%} \\
			Shinbashi 4 & NS & 49373 & 18496 & 269.80 & 12374 & 2.2{\%} \\
			Shinbashi 4 & SN & 45741 & 32013 & 249.95 & n/a & n/a \\
			Shinjuku 7 & NS & 38336 & 19473 & 209.49 & 17823 & 1.2{\%} \\
			Shinjuku 7 & SN & 36785 & 28563 & 201.01 & 16449 & 1.2{\%} \\
			\hline
		\end{tabular}
	\end{center}
\end{table*}

\begin{figure}[tbp]
	\centering
	\includegraphics[clip, width=0.99\hsize]{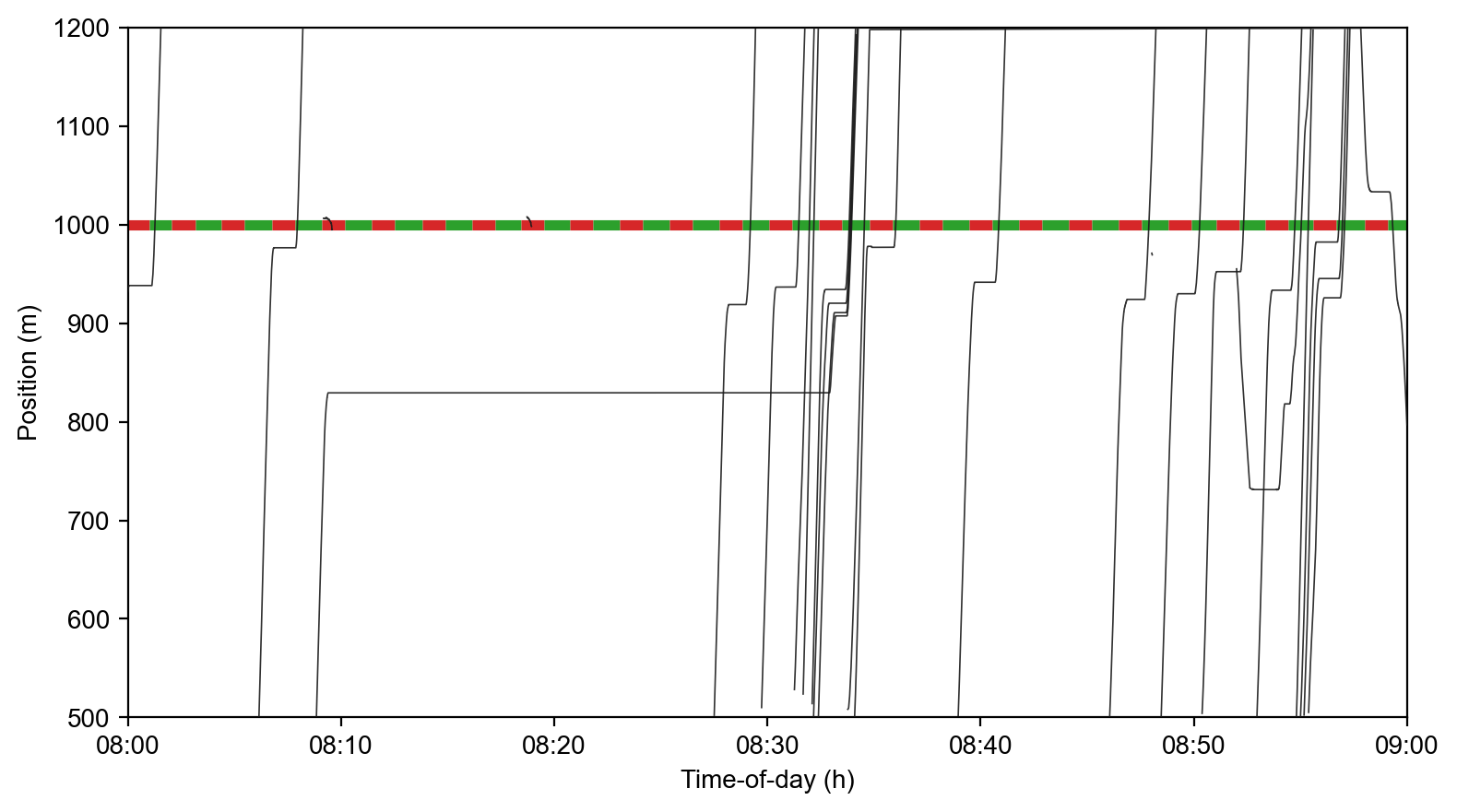}
	\caption{Sample CV trajectories at Shinbashi 4 SN on a specific day. The intersection is at position 1000 m.}
	\label{sample_CV}
\end{figure}

CV trajectories on a particular date and time are illustrated in \cref{sample_CV} as a time--space diagram.
We can confirm that CVs follow typical traffic behavior upstream of the signal, except for some anomalies.
For most cycles, the number of CVs that stop is zero or one, owing to the low penetration rate.

\begin{figure*}[tbp]
	\centering
	\subfloat[Ebisu 3 NS]{\includegraphics[clip, width=0.5\hsize]{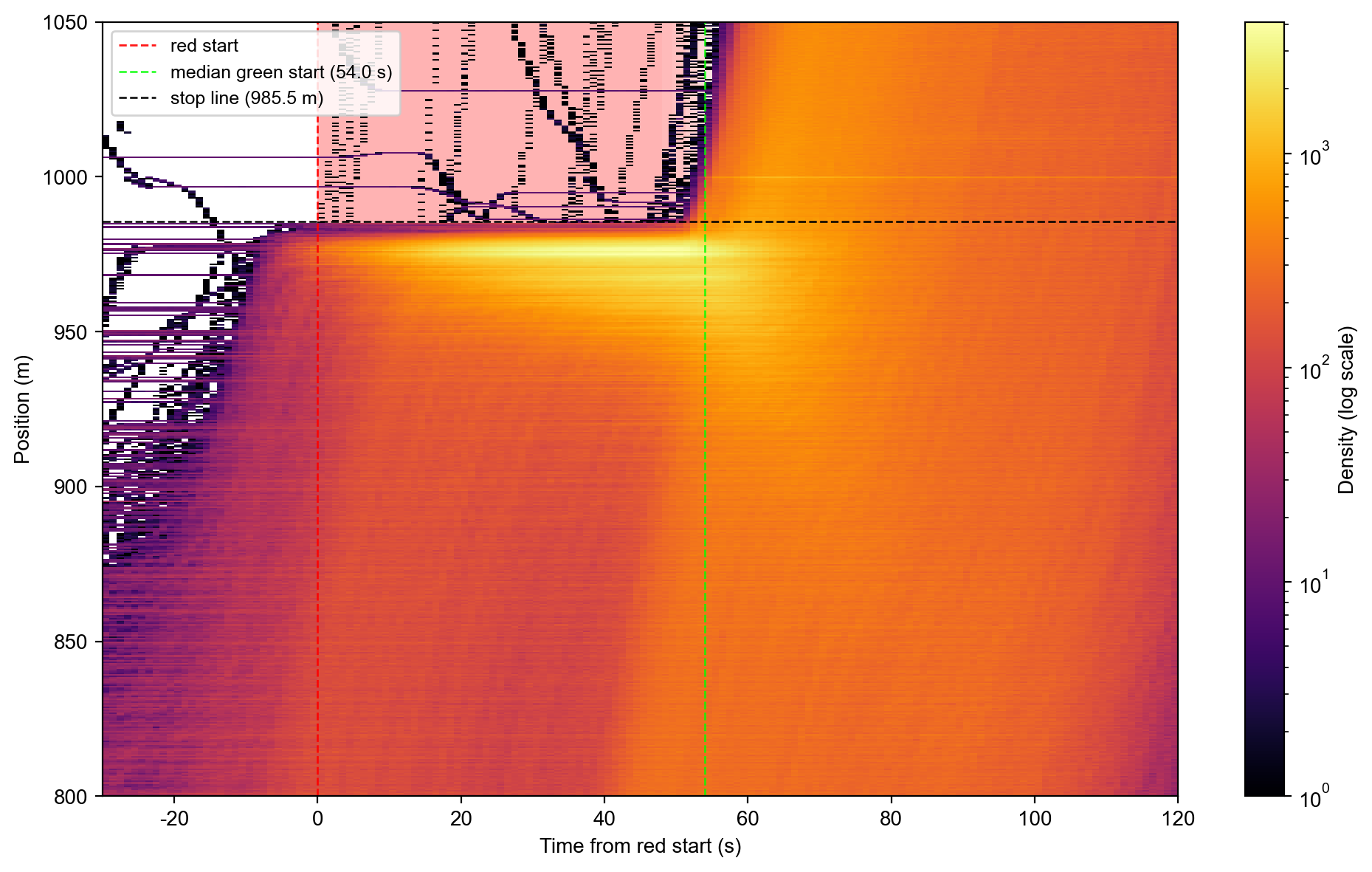}\label{traj_dist_ebis3ns}}~
	\subfloat[Ebisu 3 SN]{\includegraphics[clip, width=0.5\hsize]{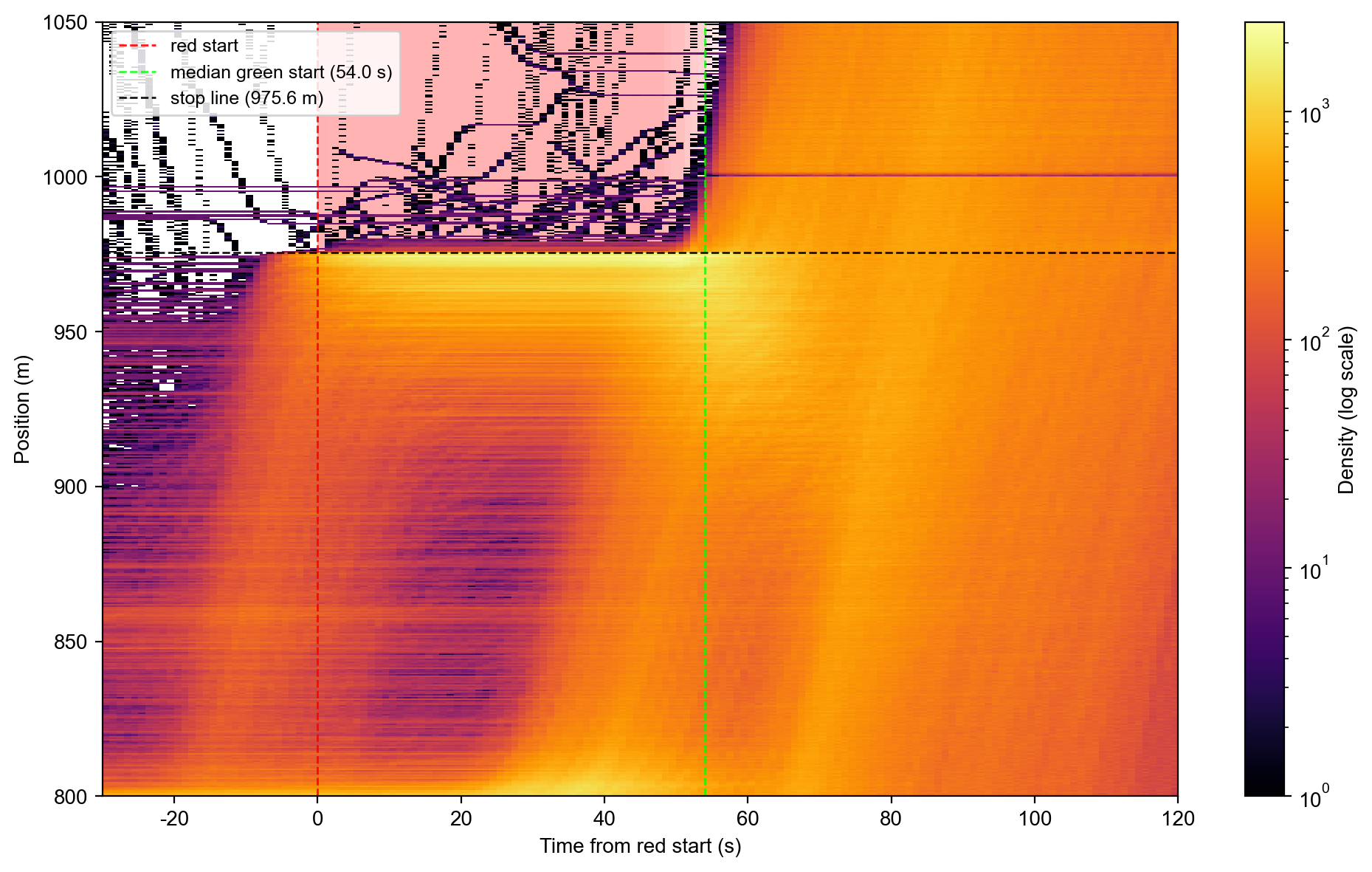}\label{traj_dist_ebis3sn}}\\
	
	\subfloat[Shinbashi 4 NS]{\includegraphics[clip, width=0.5\hsize]{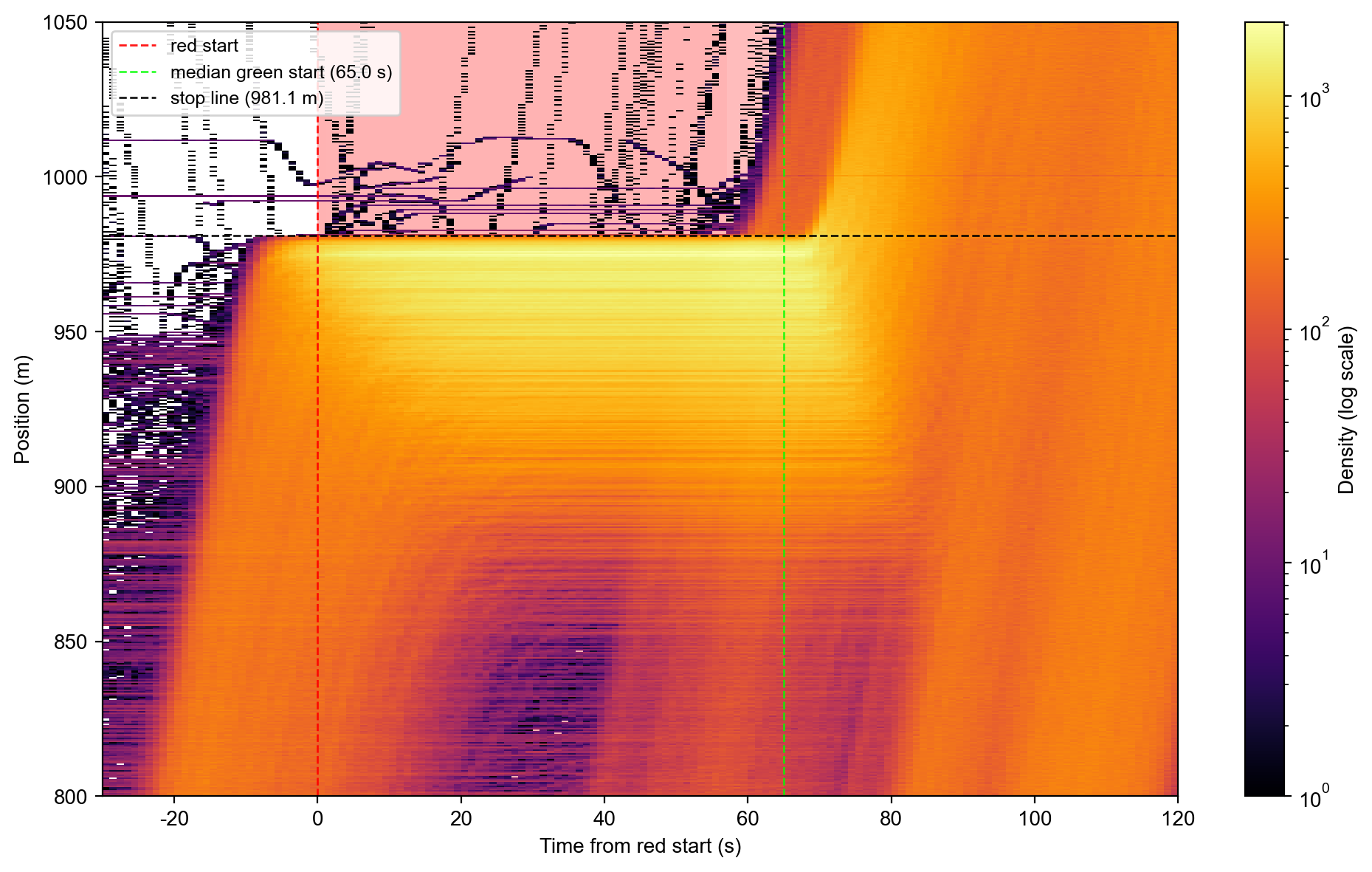}\label{traj_dist_bash4ns}}~
	\subfloat[Shinbashi 4 SN]{\includegraphics[clip, width=0.5\hsize]{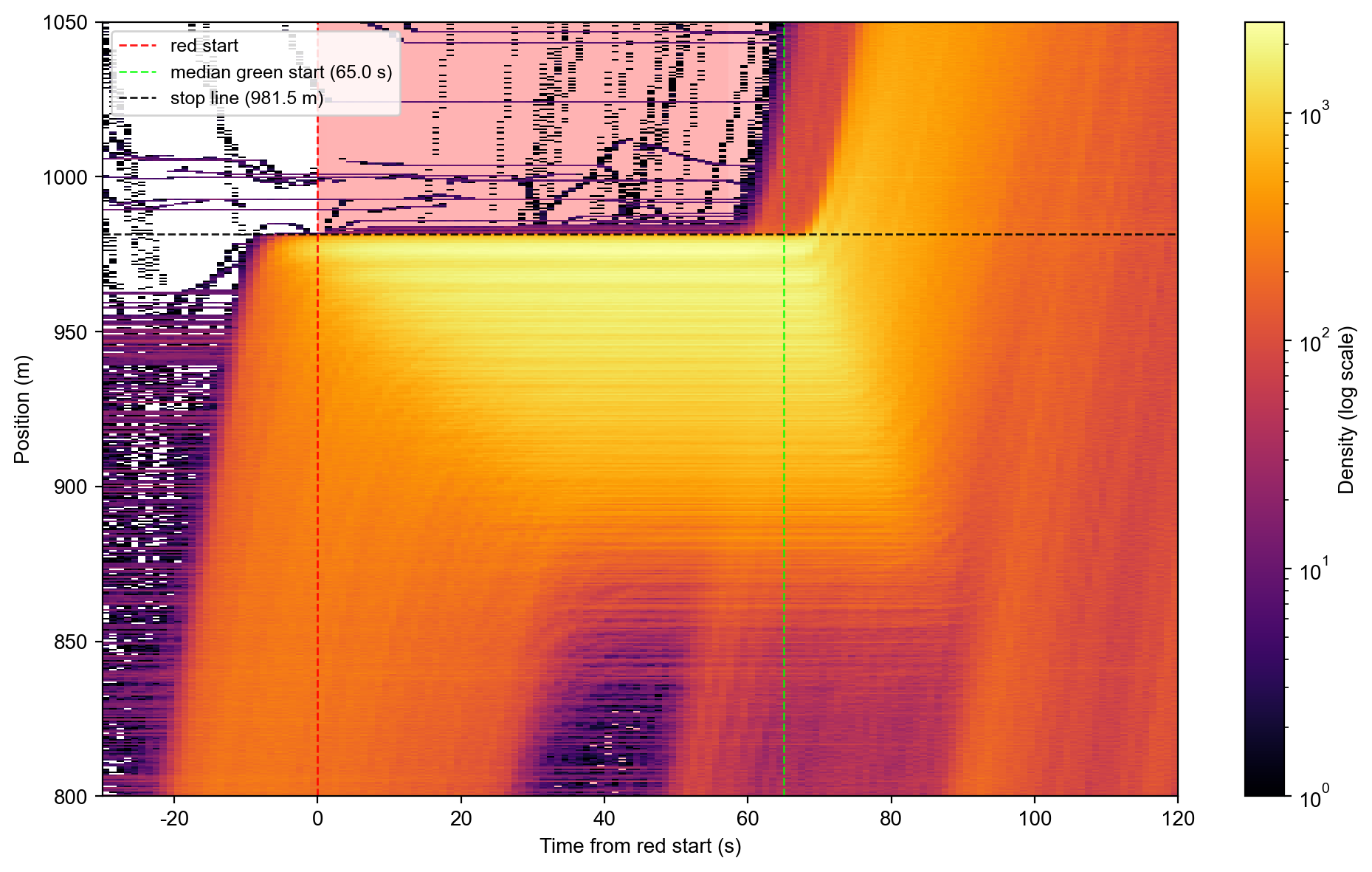}\label{traj_dist_bash4sn}}\\
	
	\subfloat[Shinjuku 7 NS]{\includegraphics[clip, width=0.5\hsize]{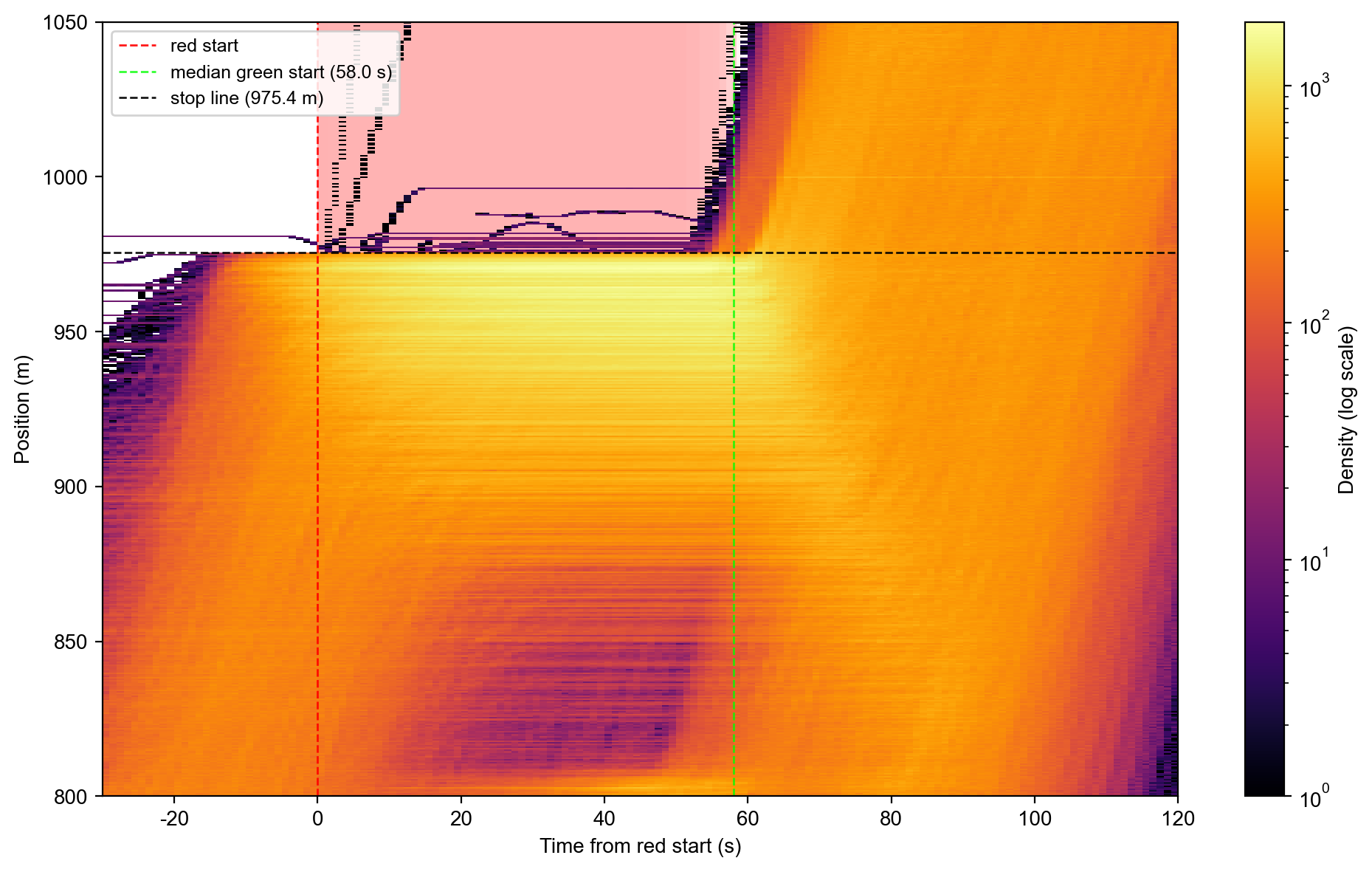}\label{traj_dist_juku7ns}}~
	\subfloat[Shinjuku 7 SN]{\includegraphics[clip, width=0.5\hsize]{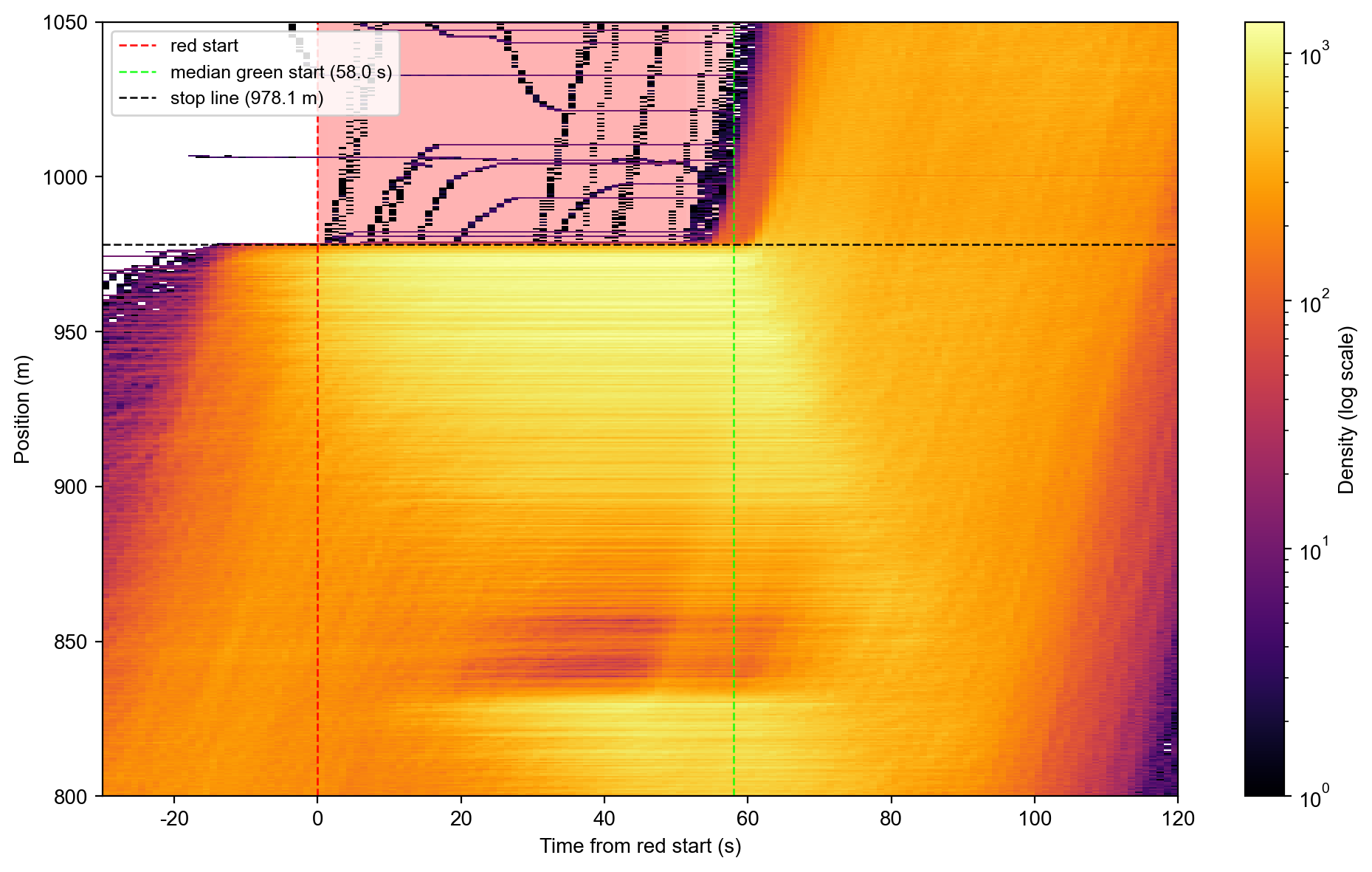}\label{traj_dist_juku7sn}}\\
	\caption{Trajectory distributions for the entire data collection duration.}
	\label{traj_distributions}
\end{figure*}

The accumulated CV trajectory distributions for the entire data collection period for each site are shown in \cref{traj_distributions}.
These trajectories are synchronized by the red start time of each cycle.
Although the penetration rate was low, we can see clear patterns by accumulating the data over a long period.
For example, two peaks just upstream of the stop line can be confirmed at most sites.
The departure and arrival waves are clearly visible.

The hyperparameters of the methods were set as follows based on common sense: $X_{\max}=30$~m, $\Delta X= 0.1$~m, $X_{\mathrm{minsat}}=20$~m, $X_{\mathrm{maxsat}}=120$~m, $T_{\mathrm{maxsat}}=120$~s.

\subsection{Results}\label{sec_results}

\begin{figure}[tbp]
	\centering
	\subfloat[Gaussian mixture method]{\includegraphics[clip, width=0.99\hsize]{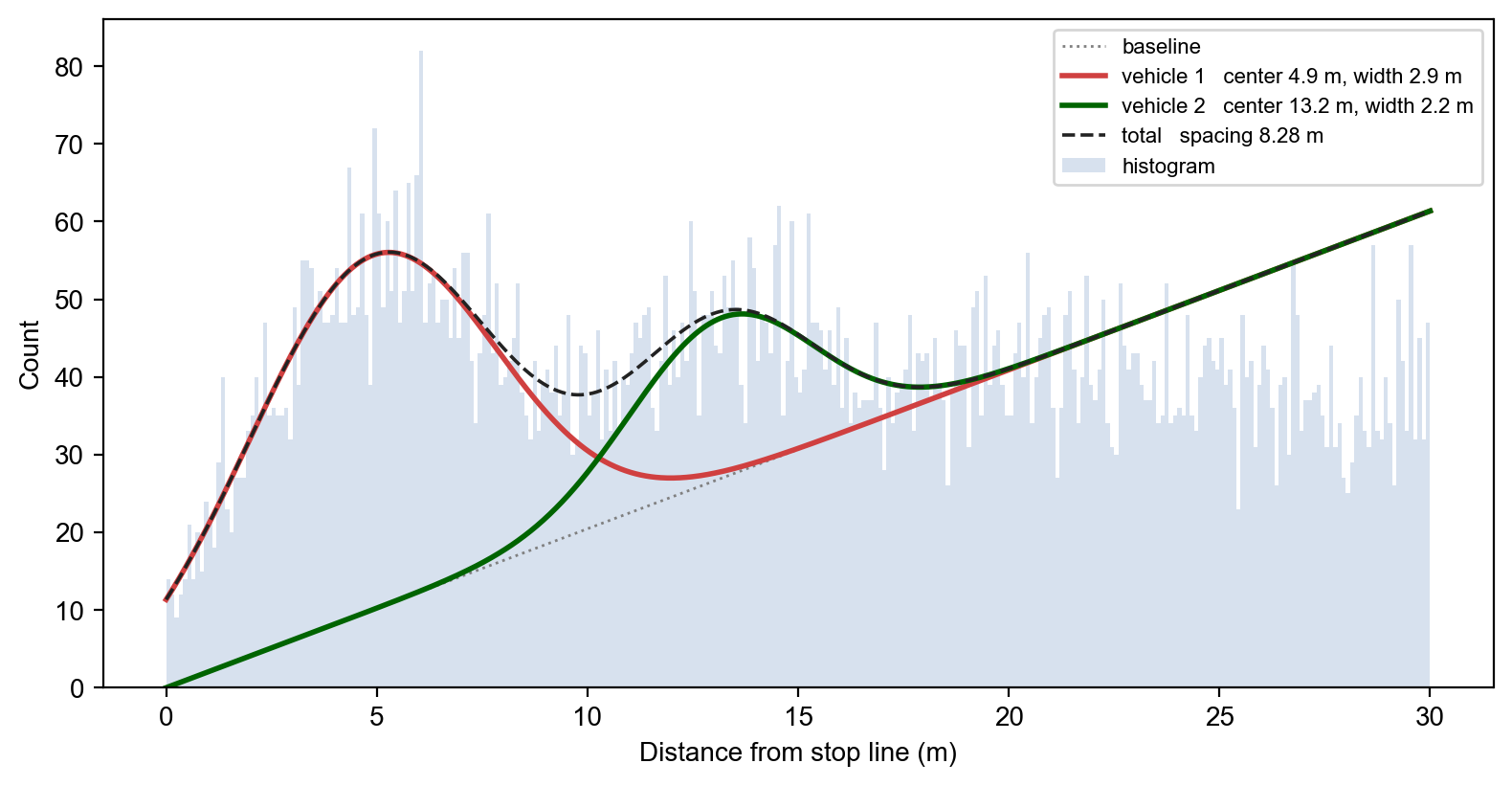}\label{jam_res_detailed_G}}\\
	\subfloat[Autocorrelation method]{\includegraphics[clip, width=0.99\hsize]{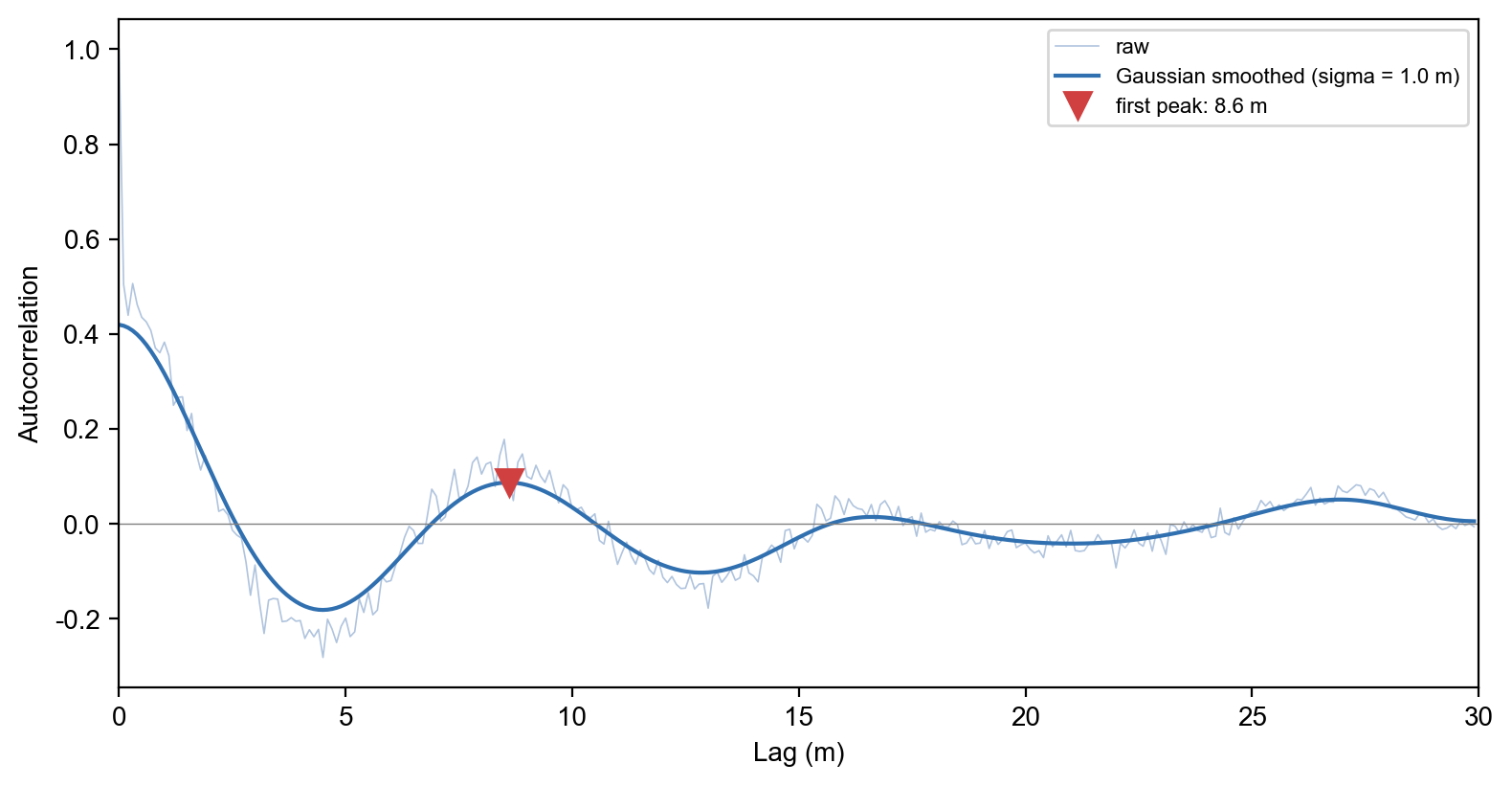}\label{jam_res_detailed_S}}\\
	\subfloat[Rule-based method]{\includegraphics[clip, width=0.99\hsize]{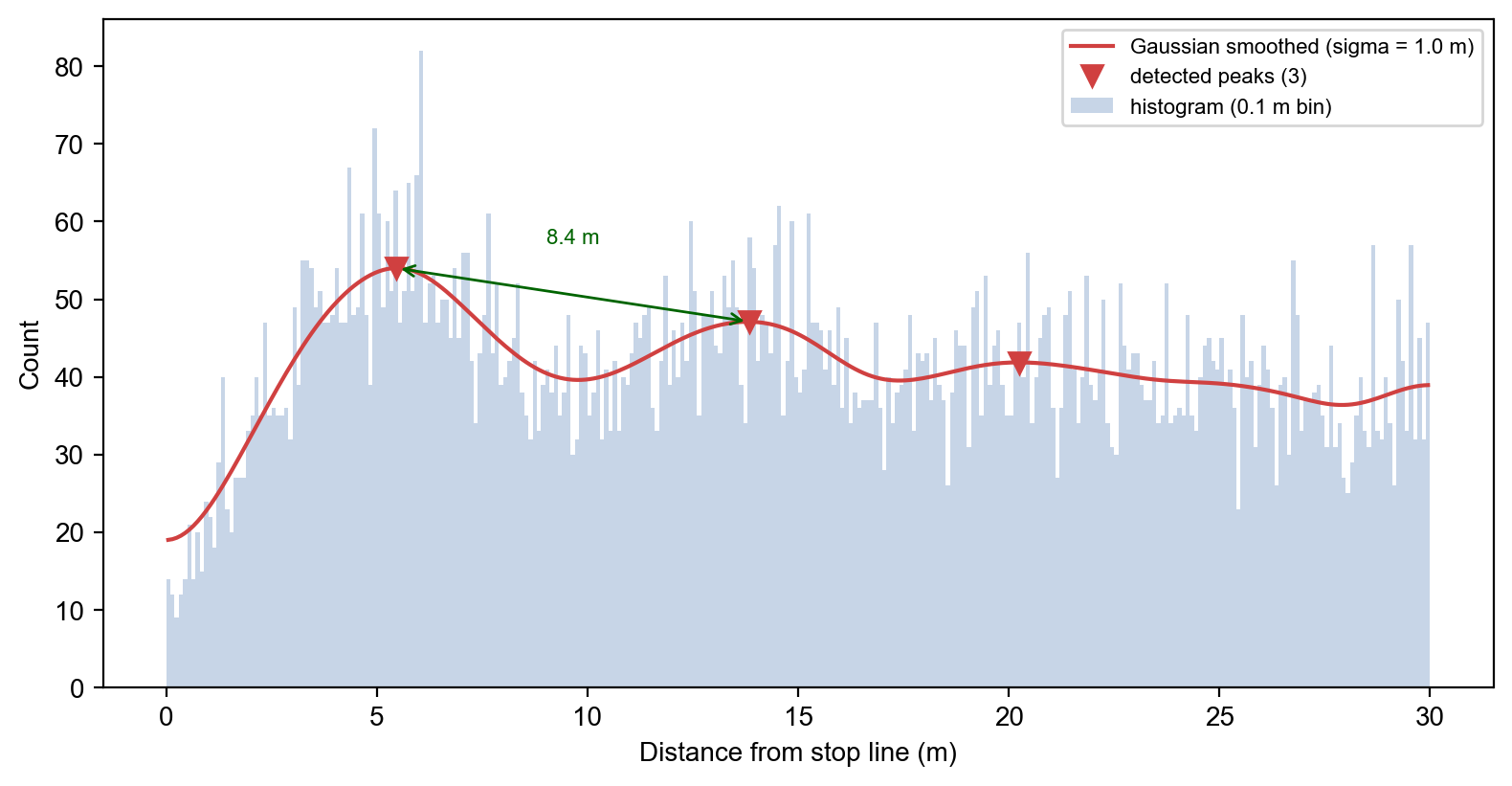}\label{jam_res_detailed_R}}\\
	\caption{The detailed results of jam spacing estimation at Shinbashi 4 SN (high accuracy site).}
	\label{jam_res_detailed}
\end{figure}

\begin{figure}[tbp]
	\centering
	\subfloat[Gaussian mixture method]{\includegraphics[clip, width=0.99\hsize]{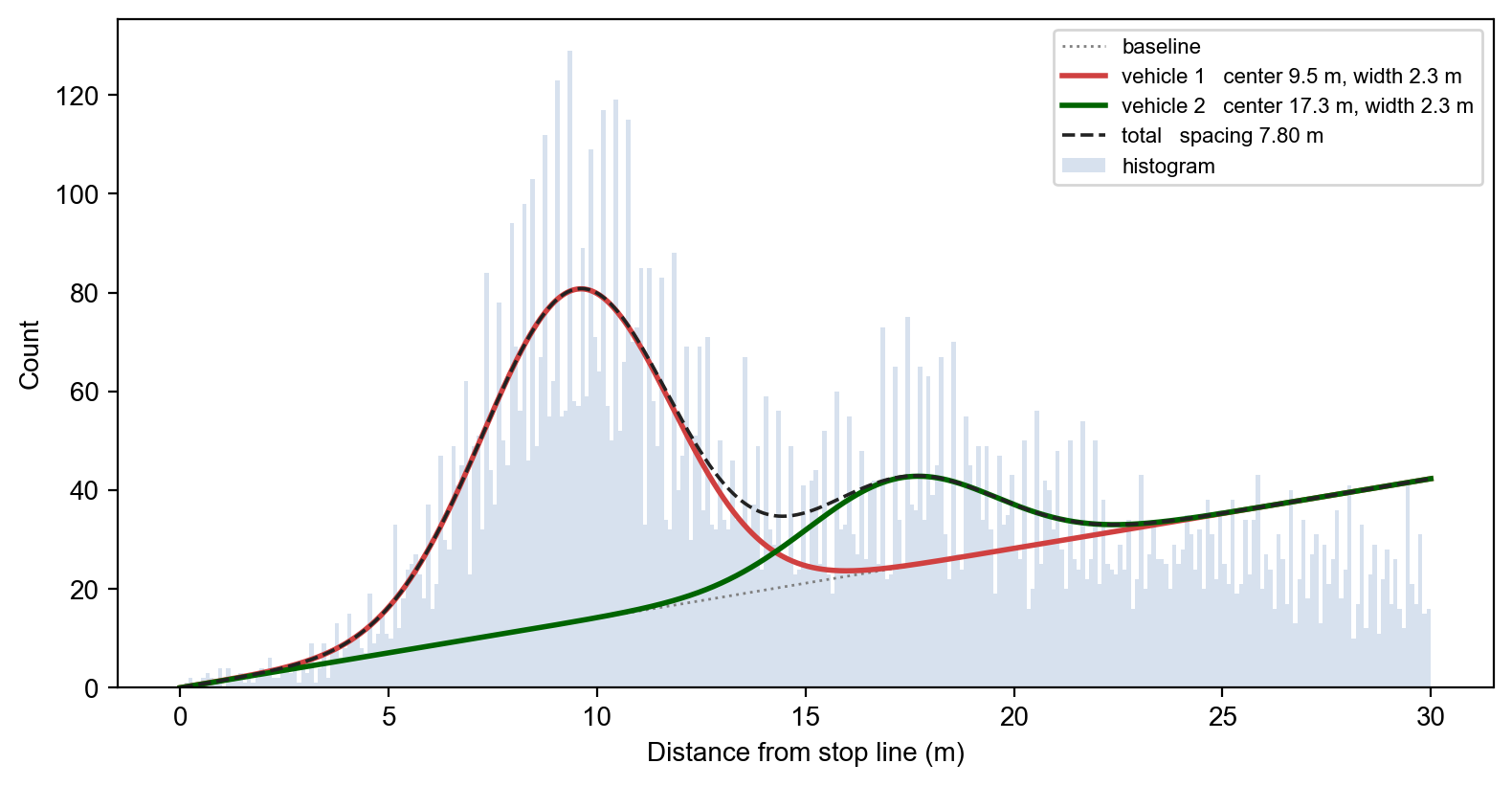}\label{jam_res_detailed_G_bad}}\\
	\subfloat[Autocorrelation method]{\includegraphics[clip, width=0.99\hsize]{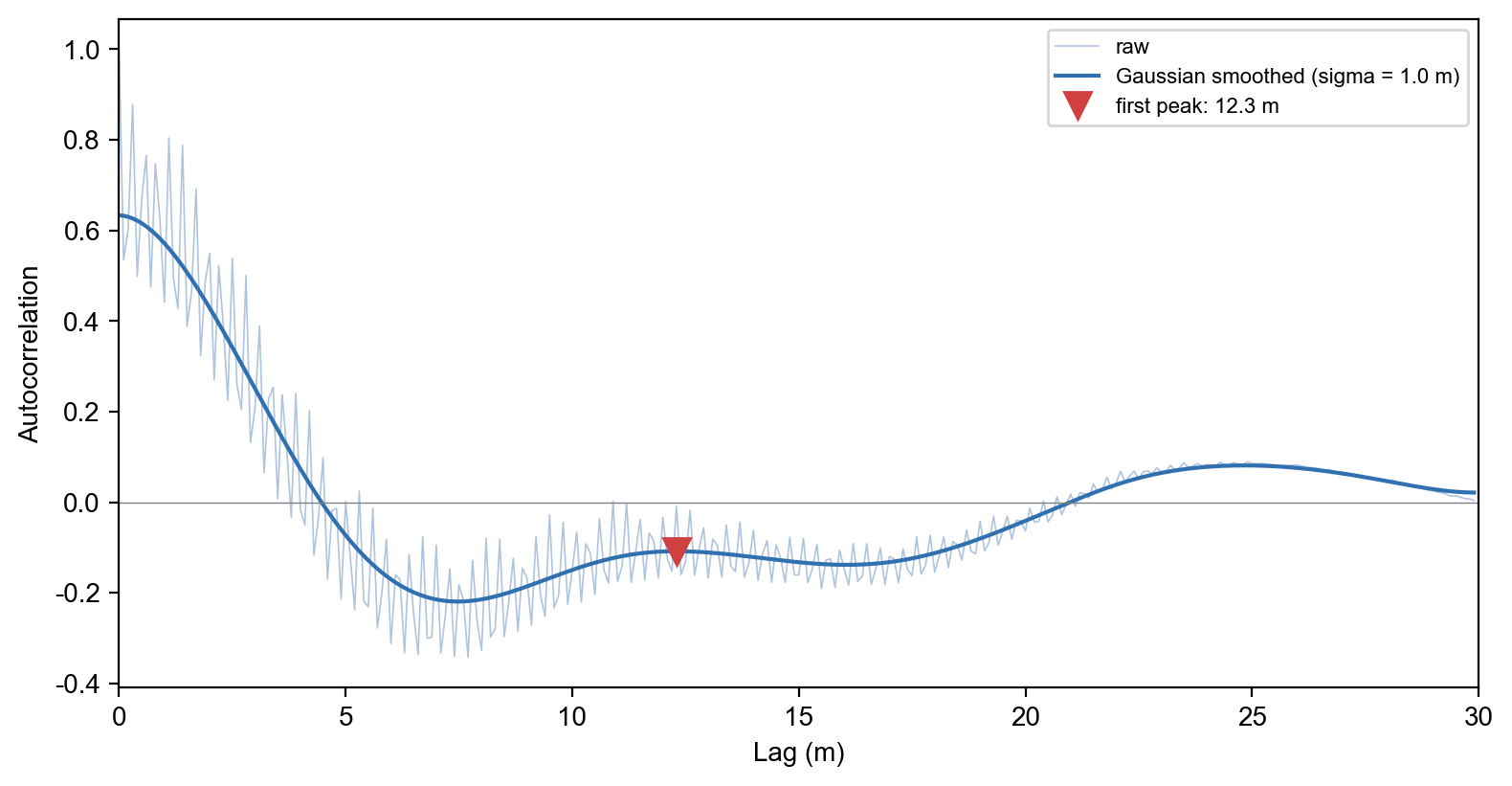}\label{jam_res_detailed_S_bad}}\\
	\subfloat[Rule-based method]{\includegraphics[clip, width=0.99\hsize]{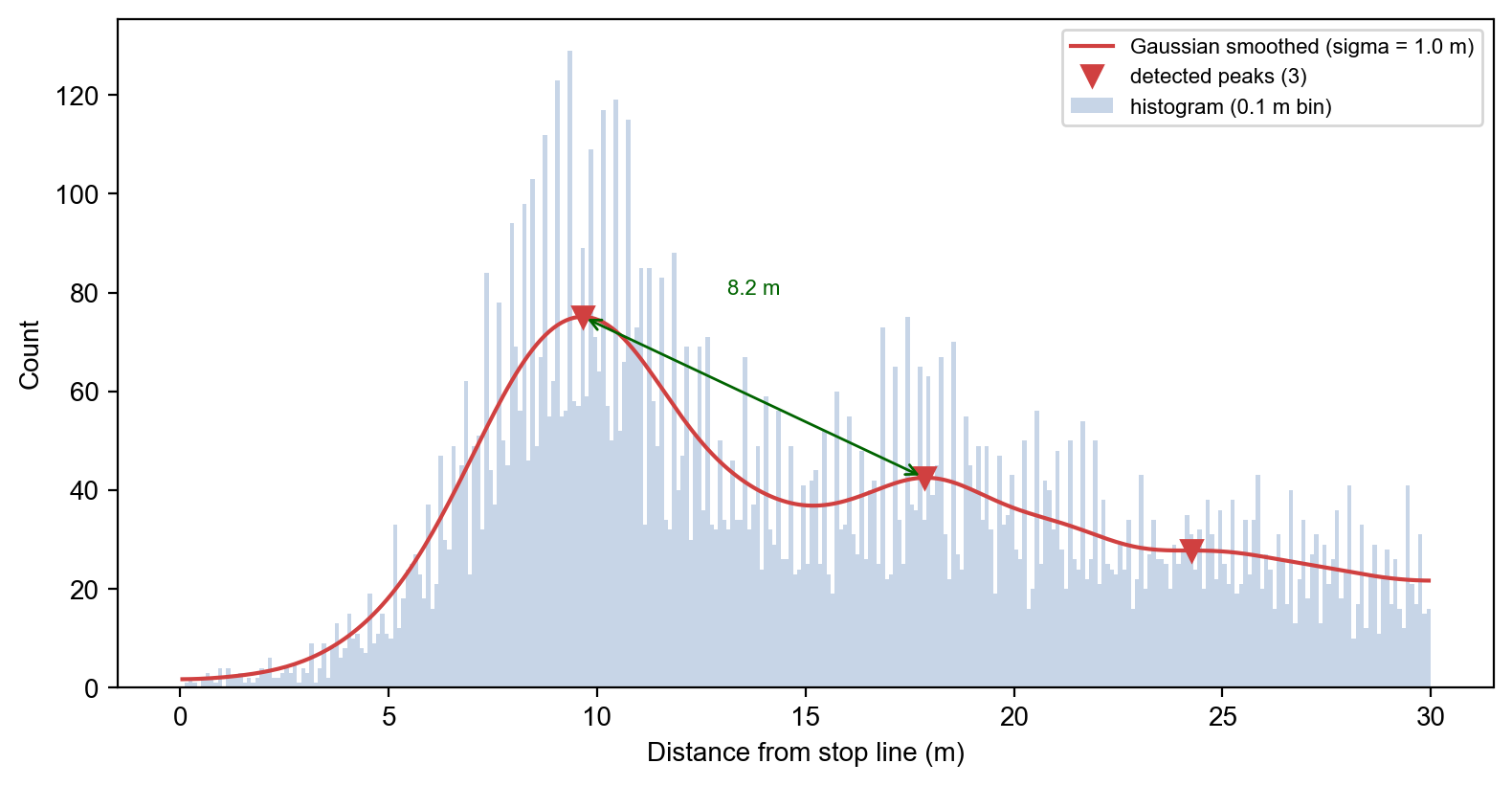}\label{jam_res_detailed_R_bad}}\\
	\caption{The detailed results of jam spacing estimation at Ebisu 3 NS (low accuracy site).}
	\label{jam_res_detailed_bad}
\end{figure}

The detailed per-method outputs at two representative sites are shown in \cref{jam_res_detailed,jam_res_detailed_bad}.
\Cref{jam_res_detailed} corresponds to Shinbashi~4 SN, which was the site at which the three methods agreed most closely with each other and with the video reference.
For this site, we can say that the data distribution is almost as expected from the theory, and consistent estimation results were obtained.
\Cref{jam_res_detailed_bad} corresponds to Ebisu~3 NS, which was the site at which the agreement among methods was the weakest.
For this site, the second peak was not very significant; as a result, the estimates were unstable, especially for the autocorrelation method.

\begin{figure}[tbp]
	\centering
	\includegraphics[clip, width=0.8\hsize]{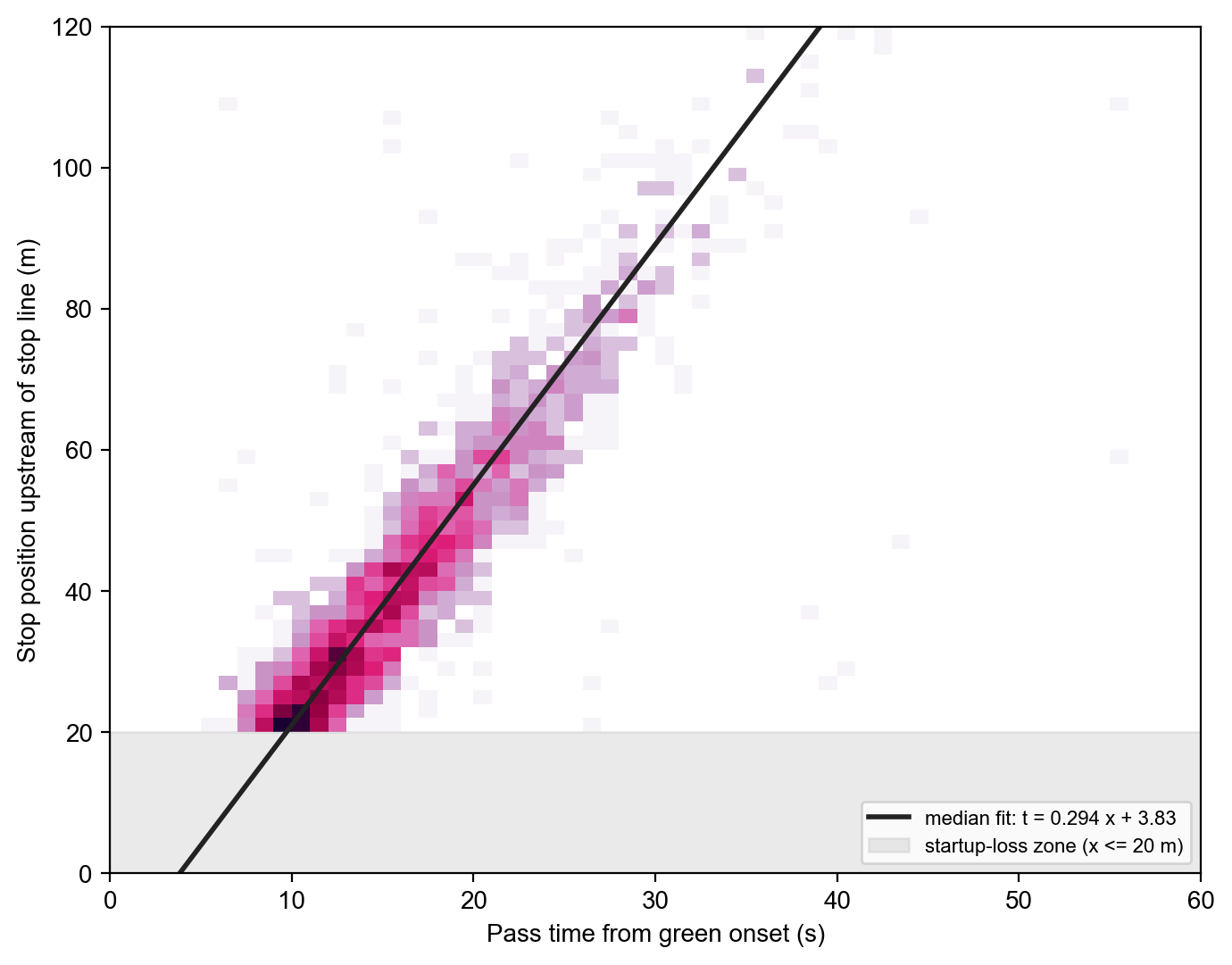}
	\caption{Distribution of the CVs' starting positions and the time elapsed from green onset.}
	\label{wave_speed_dist}
\end{figure}

Regarding the saturation flow estimation, the empirical relation between stop position and pass time from green onset at a representative site (Shinbashi 4 SN) is shown in \cref{wave_speed_dist}.
A clear pattern can be confirmed.
The resulting saturation flow estimates were approximately 1400 veh/h/lane.
The ground-truth value obtained by video recording was 1414 veh/h/lane, which is very close to the estimates.

The full estimation results are summarized in \cref{res_all}.
The ``video'' and ``MAPE'' refer to the lane-averaged saturation flow obtained from the video reference of \cite{Seo2026signalcv_en} and to the mean absolute percentage error of the estimator with respect to this reference, respectively.
The ``Gauss'', ``Autocorr'', and ``Rule'' indicate the Gauss, Autocorrelation, and Rule-based methods respectively.
Across the six sites and three methods, most jam spacing estimates lie in the range 7--9~m, which is broadly consistent with the average value of approximately 8~m reported from aerial-photo measurements in \cite{Seo2026signalcv_en}.
The corresponding capacity estimates likewise lie in the range 1200--1900~veh/h/lane; some of them are close to the ground truth, whereas some other are not.

A site-by-site comparison with the video reference reveals two main insights.
First, at the data-rich sites with regular queue formation, in particular Shinbashi~4 SN, all three methods produce capacity estimates within approximately 10{\%} of the video reference, with bootstrap standard errors below 100~veh/h/lane.
Second, at sites where the empirical histogram is less informative, the inter-method spread becomes substantially larger and individual methods can produce visibly biased estimates: the Autocorrelation method returns 12.3~m at Ebisu~3 NS (capacity error $-38{\%}$), the Gaussian mixture method returns 5.5~m at Shinjuku~7 SN (capacity error $+33{\%}$), and the Gaussian mixture method and the Rule-based method both return roughly 7~m at Shinbashi~4 NS (capacity errors of $+29{\%}$ and $+21{\%}$, respectively).
Across the six sites, the Rule-based method shows the smallest absolute percentage error overall, the Autocorrelation method is second, and the Gaussian mixture method (labeled ``Gauss'') has the largest dispersion driven by occasional fits to weak second peaks.

\begin{table*}[tbp]
	\begin{center}
		\caption{Estimation results with all CVs.}
		\label{res_all}
		\begin{tabular}{lrlrrcrrrr}
			\hline
			\multirow{2}{*}{Site} & \multirow{2}{*}{Sample} & \multirow{2}{*}{Method}
			& \multicolumn{2}{c}{jam spacing (m)}
			& & \multicolumn{4}{c}{capacity (veh/h/lane)} \\
			\cline{4-5}\cline{7-10}
			 & & & mean & std.err. & & mean & std.err. & video & MAPE \\
			\hline
			Ebisu 3 NS      & 14633 & Gauss &  7.8 & 0.3 & & 1391.8 &  54.0 & 1414 &  $-$2\% \\
			Ebisu 3 NS      & 14633 & Autocorr & 12.3 & 2.9 & &  882.4 & 104.7 & 1414 & $-$38\% \\
			Ebisu 3 NS      & 14633 & Rule &  8.2 & 0.2 & & 1323.7 &  36.4 & 1414 &  $-$6\% \\
			\hline
			Ebisu 3 SN      & 25139 & Gauss &  9.1 & 1.5 & & 1259.7 & 252.4 & 1546 & $-$19\% \\
			Ebisu 3 SN      & 25139 & Autocorr &  8.7 & 1.7 & & 1314.5 & 125.9 & 1546 & $-$15\% \\
			Ebisu 3 SN      & 25139 & Rule &  9.3 & 1.3 & & 1229.7 & 318.0 & 1546 & $-$20\% \\
			\hline
			Shinbashi 4 NS  & 18443 & Gauss &  6.7 & 1.4 & & 1876.9 & 575.5 & 1454 &  29\% \\
			Shinbashi 4 NS  & 18443 & Autocorr &  8.8 & 0.7 & & 1423.7 &  84.1 & 1454 &  $-$2\% \\
			Shinbashi 4 NS  & 18443 & Rule &  7.1 & 1.0 & & 1764.6 & 258.6 & 1454 &  21\% \\
			\hline
			Shinbashi 4 SN  & 32013 & Gauss &  8.3 & 0.3 & & 1458.6 &  51.8 & 1414 &   3\% \\
			Shinbashi 4 SN  & 32013 & Autocorr &  8.6 & 0.2 & & 1405.2 &  33.4 & 1414 &  $-$1\% \\
			Shinbashi 4 SN  & 32013 & Rule &  8.4 & 0.4 & & 1438.7 &  78.0 & 1414 &   2\% \\
			\hline
			Shinjuku 7 NS   & 19473 & Gauss &  7.3 & 1.2 & & 1405.0 & 597.0 & 1463 &  $-$4\% \\
			Shinjuku 7 NS   & 19473 & Autocorr &  7.9 & 0.4 & & 1301.6 &  55.8 & 1463 & $-$11\% \\
			Shinjuku 7 NS   & 19473 & Rule &  8.0 & 0.8 & & 1285.4 & 168.4 & 1463 & $-$12\% \\
			\hline
			Shinjuku 7 SN   & 28563 & Gauss &  5.5 & 2.4 & & 1894.3 & 1598.1 & 1423 &  33\% \\
			Shinjuku 7 SN   & 28563 & Autocorr &  8.1 & 3.3 & & 1277.1 &  273.6 & 1423 & $-$10\% \\
			Shinjuku 7 SN   & 28563 & Rule &  7.6 & 1.7 & & 1361.2 &  559.5 & 1423 &  $-$4\% \\
			\hline
		\end{tabular}
	\end{center}
\end{table*}

\subsection{Discussion}

We can draw several insights from the above results.

The first is the existence and interpretation of the peaks themselves.
Three conceptually different procedures, namely the parametric Gaussian mixture method, the spatial Autocorrelation method, and the non-parametric Rule-based method, arrive at jam spacing estimates concentrated near 8~m at the well-behaved sites.
Therefore, the peaked structure observed in the empirical stop-position histogram can be reasonably interpreted as a real signature of the underlying jam spacing, rather than as a numerical coincidence.
Although this was conjectured by the existing work \cite{LloretBatlle2023signalcv}, our analysis for this particular dataset finds that only the first two or three peaks are reliable, and the records from subsequent waiting vehicles were not informative for jam density estimation due to the accumulated noises; this can be considered as a unique findings from the actual CV data.

The second insight is about the relative behavior of the three methods.
The Rule-based method is the most robust across sites, with the smallest dispersion of percentage errors against the video reference and no individual site at which it produces a clearly biased estimate.
This robustness is likely due to the fact that the Rule-based method poses no sophisticated assumptions on the data distribution, making the method robust against unpredictable noise in real data.
The Autocorrelation method is comparably robust at most sites because it aggregates the periodic information at every lag and is therefore less dependent on any individual peak; the exception is Ebisu~3 NS, where the histogram does not exhibit a clear periodic component.
The Gaussian mixture method is efficient when the histogram closely matches the assumed two-component shape, but it can become unstable when the second peak is weak.

The third insight concerns the conditions under which the estimation succeeds.
Sites with a large number of stopped CVs, regular queue formation, and a clearly developed first-and-second-peak structure (Shinbashi~4 SN being the best example) yield agreement among methods within a few percent and capacity estimates that lie within 10{\%} of the video reference.
Sites at which queues are short or queueing behavior is otherwise irregular (Ebisu~3 NS being the clearest example) produce histograms with a weaker periodic component, and the Autocorrelation method suffers accordingly because its aggregation logic amplifies any spurious lag in the autocorrelation.
For practical deployment, the agreement among the three methods on a given site can therefore be used as a self-consistency check: a site at which all three methods return spacing values within a small range can be considered well posed, while a site at which they disagree should be flagged for closer inspection.

The fourth insight concerns the possibility of another self-consistency check via the bootstrap method.
According to \cref{res_all}, the saturation flow estimates with standard errors smaller than 100 veh/h/lane can be considered reliable, as their discrepancy from the video data is small.
This can be a useful quantitative criterion for filtering out unreliable estimates.

Meanwhile, the estimation errors might have been significantly influenced by factors that are not considered by the methods, that is, inter-lane heterogeneity.
For example, at Shinbashi 4 NS, a relatively large amount of curb-side parking is commonly observed.
At Ebisu 3 SN and Shinjuku 7 SN, bus stops are located nearby.
The downstream road of Shinjuku 7 SN has a bus-only lane during the peak periods.
These facts may have introduced inter-lane heterogeneity, which caused the relatively large errors at these sites by making it difficult to apply the proposed methods, which assume that vehicle stop positions are homogeneous across all lanes.

\section{Conclusion}

This paper investigated methods for estimating the jam density and the saturation flow at signalized intersections from low penetration rate CV trajectory data alone.
The methods exploit the long-term statistical pattern of CV stop positions accumulated over many signal cycles, and combine peak-based jam spacing estimators with the regression-based capacity estimator.
In order to obtain realistic insights, the methods were evaluated by applying them to actual CV data collected over six months at six different sites in Tokyo.
The proposed methods themselves are similar to one proposed by an existing theoretical work \cite{LloretBatlle2023signalcv}; however, findings from in-depth validation using actual CV data can be considered as unique and helpful for actual implementation.

The empirical results yields the following conclusions.
The peaked structure expected from queueing geometry is observed consistently in the empirical CV stop-position histogram, and the three methods agree on the resulting jam spacing scale at the sites where this structure is well developed.
At sites with well-formed queues, the resulting capacity estimates lie within roughly 10{\%} of an independent video-based reference.
We revealed that the methods work reasonably under certain conditions, and propose a quantitative criteria for self-consistency check: agreement among the three methods and standard error less than 100 veh/h/lane.
These insights would be useful for implementing jam density and saturation flow estimation methods, as well as traffic signal control methods based on these estimated parameters. 
This would lead to CV-driven traffic management that is applicable even with sparse CV availability.

\section*{Acknowledgements}

This work was conducted as a part of the ``Committee on novel traffic control methods'' of the Japan Society of Traffic Engineers, funded by the UTMS Society of Japan.
The CV data were provided by UTMS Society of Japan.
We received valuable comments from the committee participants. 
We express our sincere appreciation to all involved.

\section*{Conflict of interest}

The authors declare that they have no known conflict of interest.

\bibliographystyle{spmpsci}      

\end{document}